\documentclass[pdflatex,sn-nature]{sn-jnl}

\usepackage{graphicx}%
\usepackage{multirow}
\usepackage{amsmath,amssymb,amsfonts}%
\usepackage{amsthm}%
\usepackage{mathrsfs}%
\usepackage[title]{appendix}%
\usepackage{xcolor}%
\usepackage{textcomp}%
\usepackage{booktabs}%
\usepackage{algorithm}%
\usepackage{algorithmicx}%
\usepackage{algpseudocode}%
\usepackage{listings}%
\usepackage{hyperref}%
\usepackage{natbib}
\usepackage{graphicx}
\usepackage{subcaption}
\usepackage{longtable}
\usepackage{float} 
\usepackage{textgreek} 
\usepackage{lscape} 
\usepackage{placeins}
\usepackage{comment}
\usepackage{rotating}
\usepackage{graphicx}
\usepackage{subcaption}

\theoremstyle{thmstyleone}%
\theoremstyle{thmstyletwo}%

\theoremstyle{thmstylethree}%

\usepackage{geometry}
\begin{document}

\title[Subcortical Shape Changes in the Elderly]{Subcortical Shape Variations and Their Associations with Cognition Across the 8th Decade of Life. A Study in the Lothian Birth Cohort 1936}


\author[1]{\fnm{Maria del C.} \sur{Valdes-Hernandez}}\email{M.Valdes-Hernan@ed.ac.uk}
\equalcont{These authors contributed equally to this work.}

\author*[2]{\fnm{Wonjung} \sur{Park}}\email{fabiola@kaist.ac.kr}
\equalcont{These authors contributed equally to this work.}

\author[3,4]{\fnm{Joanna} \sur{Moodie}}\email{joanna.moodie@ed.ac.uk}

\author[1]{\fnm{Susana} \sur{Muñoz Maniega}}\email{s.m.maniega@ed.ac.uk}

\author[3,4]{\fnm{Janie} \sur{Corley}}\email{Janie.Corley@ed.ac.uk}

\author[1]{\fnm{Fraser N.} \sur{Sneden}}\email{fsneden@hotmail.com}

\author[1]{\fnm{Mark E.} \sur{Bastin}}\email{Mark.Bastin@ed.ac.uk}

\author[1]{\fnm{Joanna M.} \sur{Wardlaw}}\email{Joanna.Wardlaw@ed.ac.uk}

\author[3,4]{\fnm{Simon R.} \sur{Cox}}\email{Simon.Cox@ed.ac.uk}

\author*[2]{\fnm{Jinah} \sur{Park}}\email{jinahpark@kaist.ac.kr}

\affil[1]{\orgdiv{Department of Neuroimaging Sciences}, \orgname{University of Edinburgh}, \orgaddress{\street{49 Little France Crescent}, \city{Edinburgh}, \postcode{EH164SB}, \country{United Kingdom}}}

\affil[2]{\orgdiv{Computer Graphics and Visualization Laboratory}, \orgname{Korea Advanced Institute of Science and Technology (KAIST)}, \orgaddress{\street{291 Daehak-ro}, \city{Daejeon}, \postcode{34141}, \country{Republic of Korea}}}

\affil[3]{\orgdiv{Department of Psychology}, \orgname{University of Edinburgh}, \orgaddress{\street{7 George Square}, \city{Edinburgh}, \postcode{EH89JZ}, \country{United Kingdom}}}

\affil[4]{\orgdiv{Edinburgh Futures Institute}, \orgname{University of Edinburgh}, \orgaddress{\street{1 Lauriston Place}, \city{Edinburgh}, \postcode{EH39EF}, \country{United Kingdom}}}


\abstract{The study of brain morphology changes in normal individuals may capture aspects of functionally-relevant brain aging not fully indicated by gross volumetry. Despite the important role of subcortical brain structures in cognition, the associations between their morphological trajectories and cognitive changes in aging have not been documented. We use neuroimaging, demographic, and cognitive data from a large longitudinal study of cognitive aging, the Lothian Birth Cohort 1936, to explore shape changes in subcortical brain structures of community-dwelling individuals across their 8th decade of life. We investigate the association of these changes with cognitive aging using ANCOVA and mixed linear model analyses. Subcortical shape changes were heterogeneous, with varied atrophy patterns across whole period. The hippocampus and the ventral DC experienced varied morphological deformations (from its baseline point) different in left and right hemispheres, while the thalami and globus pallidi shapes, for example, experienced a more uniform volume contraction, nearly symmetrical throughout different timelines. Changes in general cognition were mainly associated with inwards and outwards vertex displacements between the time-points.}

\keywords{subcortical brain structures, hippocampus, thalamus, caudate, putamen, globus pallidus, nucleus accumbens, ventral diencephalon, shape model, shape deformations, cognition}



\maketitle

\section{Introduction}\label{sec:introduction}

The morphology of the brain and its structures is underpinned by their cytoarchitectural and hodological properties, which fluctuate by regions within these complex structures and whose implications in human behavior cannot be explained or described by volumetry alone. The study of changes in brain morphology is crucial to understand brain development and aging processes \cite{thompson2000growth}, to discern how different factors, including heritability \cite{roshchupkin2016heritability} \cite{ge2016multidimensional}, shape the brain, and for studying different diseases \cite{qiao2025subcortical}. Even among healthy individuals, there is a high degree of variability in the appearance of brain structures \cite{roshchupkin2016heritability}. Characterizing this wide range of natural differences has been pivotal in distinguishing typical variations within a "normal" range from abnormal changes associated with disease \cite{zhu2025identifying} \cite{chen2015prospective} \cite{laansma2024worldwide} or other conditions \cite{kim2020comparison} \cite{qiao2025subcortical}. It also has shed light on the mechanisms behind lifetime changes in cognitive abilities, behavioral patterns, and clinical conditions or disease progression. 

Various studies have mapped brain morphology trajectories in normal individuals across different age ranges. For example, a study in children aged 9 to 16 years mapped the trajectories of thalamic volumes and variations in cortical thickness throughout 13-16 years to explore the relationship between thalamic volume, cortical maturation, and the course of obsessive-compulsive symptoms \cite{weeland2024longitudinal}. Other studies have mapped global and regional cortical morphometric phenotypes and volumes of brain tissues throughout life in normal individuals \cite{bethlehem2022brain} and created four-dimensional quantitative maps of up to four-year growth patterns in the developing human brain of children aged 3 to 15 years \cite{thompson2000growth}. But, to our knowledge, the subcortical shape trajectories in cognitively normal elderly individuals and their association with their general cognitive abilities throughout the same life period have not yet been documented.

The lack of documentation in the patterns of the morphological brain changes in these structures with age in relation to cognition is not due to lack of knowledge about their involvement in cognitive functions. The thalamus is known to facilitate communication between the prefrontal cortex and other brain regions, integrate information for executive functions (e.g., planning and problem-solving), intervene in attention processes, and encode and retrieve memories \cite{saalmann2015cognitive} \cite{wolff2019cognitive}. The caudate plays an important role in procedural learning, associative learning, and in inhibitory control of actions \cite{grahn2008cognitive}. The putamen is involved in learning and motor control, including speech articulation, language functions \cite{vinas2017role}, reward-related processes, and addiction \cite{haber2016corticostriatal}. The globus pallidus and the hippocampus are involved in memory processes \cite{gillies2017cognitive}. A study in elderly cognitively normal individuals and patients with mild cognitive impairment at high risk of Alzheimer's disease progression found that the volume of the normal-appearing grey matter in the globus pallidus was associated with the outcome of the short-term memory binding test in the patients, suggesting support of the globus pallidus to memory binding \cite{hernandez2020striatum}. The roles of the amygdala in memory, emotional learning, and social cognition  \cite{DOMINGUEZBORRAS2022359} have long been recognised. The nucleus accumbens has been considered an interface between cognition, emotion, and action \cite{floresco2015nucleus}, moderating the later and  promoting motivational behaviour. The ventral diencephalon is a complex structure reported to be involved mainly in attention processes \cite{vossel2014dorsal}. Despite gross total volumetric indices of these structures have shown associations with function and clinical-behavioral phenotypes, studying their morphology might offer a more nuanced understanding of the anatomy of clinical conditions and normal aging trajectories. Specifically, understanding their morphological changes in relation to general cognition in a community-dwelling cohort, will put us a step closer to a precise mapping of the overall morphological change of the brain with age.

We use neuroimaging, demographic, and cognitive data from a longitudinal study of cognitive aging, The Lothian Birth Cohort 1936 (LBC1936) Study, to explore and document shape changes in subcortical brain structures of community-dwelling individuals in their 8th decade of life and investigate the association of these changes with their general cognition. We hypothesize that 1) subcortical shape changes are heterogeneous, with varied atrophy patterns across the 8th decade of life, i.e., some regions at some point looking enlarged while others shrink, 2) overall, there will be a shrinkage reflected in a reduction in volume although its degree will not be uniform throughout the decade, and 3) changes in general cognition will be associated with regions of shape change.

\section{Results}\label{sec:results}

\subsection{Sample characteristics}

Recruitment and attrition figures in the LBC1936 Study up to wave 4 have been previously published \cite{aribisala2023sleep}. From the first wave involving a brain magnetic resonance imaging (MRI) scan (wave 2) to wave 5, 866 (wave 2, mean age 72.5 $\pm$ 0.71 years), 697 (wave 3, mean age 76.24 $\pm$ 0.68 years), 550 (wave 4, mean age 79.32 $\pm$ 0.62 years) and 431 (wave 5, mean age 82.0 $\pm$ 0.47 years) participants undertook the cognitive tests (\autoref{fig:recruitment}). In the first scanning wave 700 study participants initiated the MRI session, in the second 493, in the third 390, and in the fourth 309.

\subsubsection{Subcortical structures - volumetric trajectories}

\begin{figure}[H]
\centering
\includegraphics[width=1\textwidth]{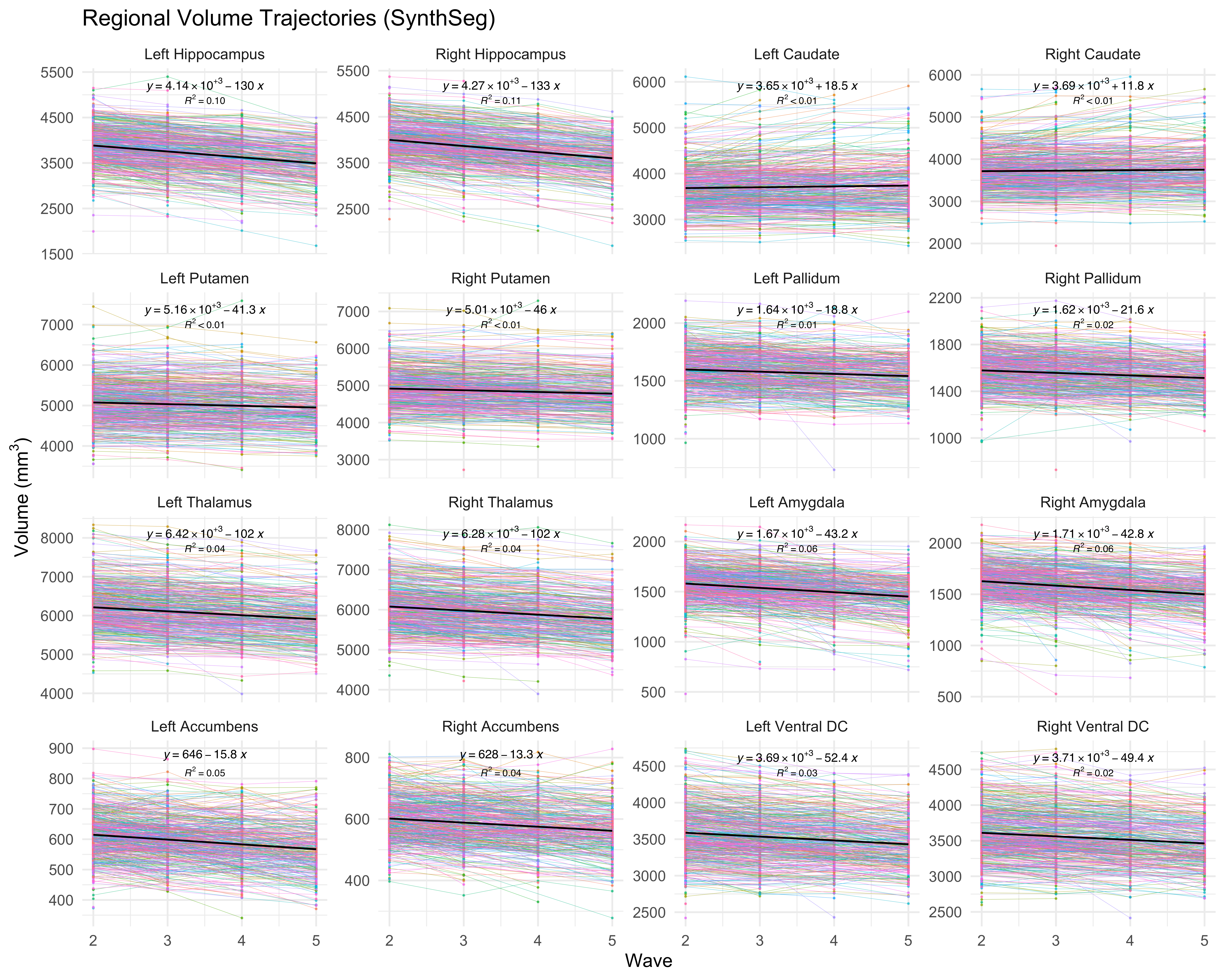}
\caption{Volumetric trajectories of each subcortical structure.}\label{fig:volumes}
\end{figure}

For this Scottish cohort, the volumetric trajectories of each of the subcortical structures between 72.5 (wave 2) and 82 (wave 5) years old, are shown in \autoref{fig:volumes}. The left and right individual trajectories are not equal, nor the general slopes with the exception of the thalami, where the trend is for the volume to decrease 102 mm$^3$ every three years for both: left and right. The caudate has positive general slopes, with the left caudate being more steep than the right, i.e., 18.5 mm$^3$ vs. 11.8 mm$^3$ volume increase every 3 years. For the rest of the subcortical structures, the average volumetric decrease with time is more pronounced in the right hemisphere than in the left (\autoref{fig:volumes}). General trends in volume changes, however, do not reflect individual trajectories (see R$^2$ values in \autoref{fig:volumes}, all $<=$0.1).

\subsubsection{Cognition and brain health - trajectories}

\begin{figure}[H]
\centering
\includegraphics[width=1\textwidth]{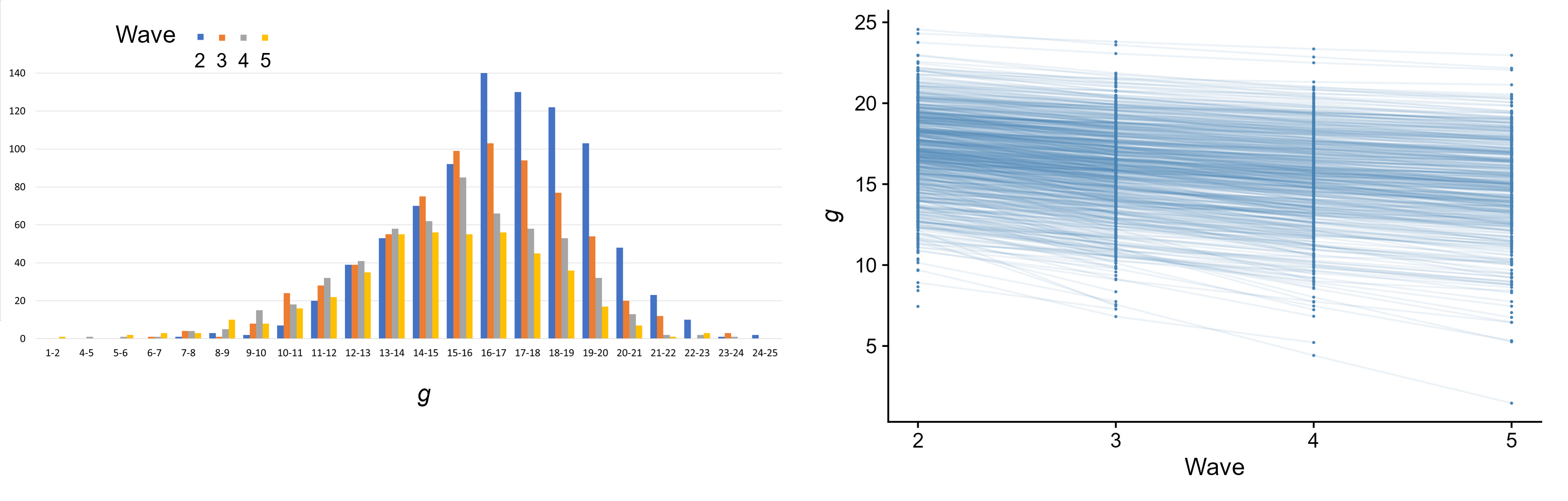}
\caption{Histograms of \textit{g} at cognitive testing waves 2 to 5 (left) and individual trajectories across the four waves of testing (right). \textit{g} values are scaled to the Block Design test, as it was the reference indicator in the model used to derive them. }\label{fig:cognition}
\end{figure}

The distribution of the values of general cognition (\textit{g}) across the cognitive testing waves, scaled to the Block Design test (reference indicator in the latent growth curve model that generated the values for this variable), is shown in \autoref{fig:cognition}. Whilst the majority of participants have \textit{g} values in the range of 15 to 20 at wave 2 (scan 1), at wave 5 (scan 4) \textit{g} values range mainly from 13 to 17, indicating a general trend of cognitive decline with increasing age. The contributions from each cognitive test to this overall cognitive score \textit{g} are in \autoref{tab:tests_loadings_table}. 

\autoref{tab:Descriptives} shows the mean and standard deviation of the values of \textit{g}, brain health index (BHI; \cite{dickie2018brain}) and brain parenchymal fraction (BPF; i.e., percentage of brain tissue volume in the intracranial vault) in each wave. The difference between the BPF and the BHI expressed as percentage of its maximum value (i.e., 1) increases with time as the number and extent of vascular abnormalities also increases -  contribution that is reflected in the BHI and not in the BPF. Although the distribution of the data that contributed for the analyses in each wave did not differ across waves (i.e., normal distribution, see \autoref{fig:cognition}), the means were all statistically significantly different. Results from the repeated ANOVA analyses can be seen in the accompanying data repository.

\begin{longtable}[c]{cccc}
\caption{Cognitive and brain health indicators in the sample that contributed to the cross-sectional analyses of the association between \textit{g} and subcortical shape variations at each wave.}
\label{tab:Descriptives}\\
\toprule[.8mm]
          & wave no. & n   & Mean (SD)        \\
          \cmidrule{1-4}
\endfirsthead
\multicolumn{4}{c}%
{{\bfseries Table \thetable\ continued from previous page}} \\
\toprule[.8mm]
          & wave no. & n   & Mean (SD)        \\
          \cmidrule{1-4}
\endhead
\multirow{4}{*}{g}         & 2        & 866 & 16.891 (2.584)   \\
          & 3        & 697 & 15.946 (2.765)   \\
          & 4        & 550 & 15.248 (2.936)   \\
          & 5        & 431 & 14.903 (3.031)   \\
          \cmidrule{1-4}
age 11 IQ &          & 816 & 100.676 (15.294) \\
\cmidrule{1-4}
\multirow{4}{*}{BHI}       & 2        & 663 & 0.681 (0.0240)   \\
          & 3        & 482 & 0.666 (0.0256)   \\
          & 4        & 383 & 0.653 (0.0456)   \\
          & 5        & 301 & 0.643 (0.0437)   \\
          \cmidrule{1-4}
\multirow{4}{*}{BPF}      & 2        & 662 & 68.967 (2.240)   \\
          & 3        & 482 & 67.737 (2.311)   \\
          & 4        & 383 & 67.0984 (2.406)  \\
          & 5        & 301 & 66.263 (2.191) \\
\bottomrule[.8mm] 
\end{longtable}

In the imaged sample (at wave 2, n=700), the sex split was 371/329 (M/F). From self-reported questionnaire, 48.7\% were hypertensive, 11\% were diabetic, 41.4\% had hypercholesterolaemia, and 27.2\% had a previous history of cerebrovascular disease. Not all participants who underwent cognitive testing (Tested) at each wave or had a brain MRI scan at the first scanning wave (i.e., wave 2, Imaged), or even initiated the MRI scan at each wave, contributed with valid data to the longitudinal analyses of changes between two assessment waves or to the cross-sectional analyses due to the reasons described in \autoref{fig:recruitment}. \autoref{tab:TAB_Missing_values} shows the results of comparing the cognitive, clinical, demographic and imaging parameters in the group of individuals  that contributed data to these analyses (Contrib) versus the overall sample that underwent cognitive testing (Tested) at each wave, and the sample that did not contribute (Missed). Although these groups of individuals did not differ in terms of visually assessed superficial brain atrophy, those who did not contribute to the analyses at waves 4 and 5 differed from the rest in all the remaining tabulated parameters. Age 11 IQ differed throughout between contributors and not contributors.


\subsection{Associations between shape morphology changes and cognitive abilities}
\begin{figure}[b]
\centering
\includegraphics[width=0.85\textwidth]{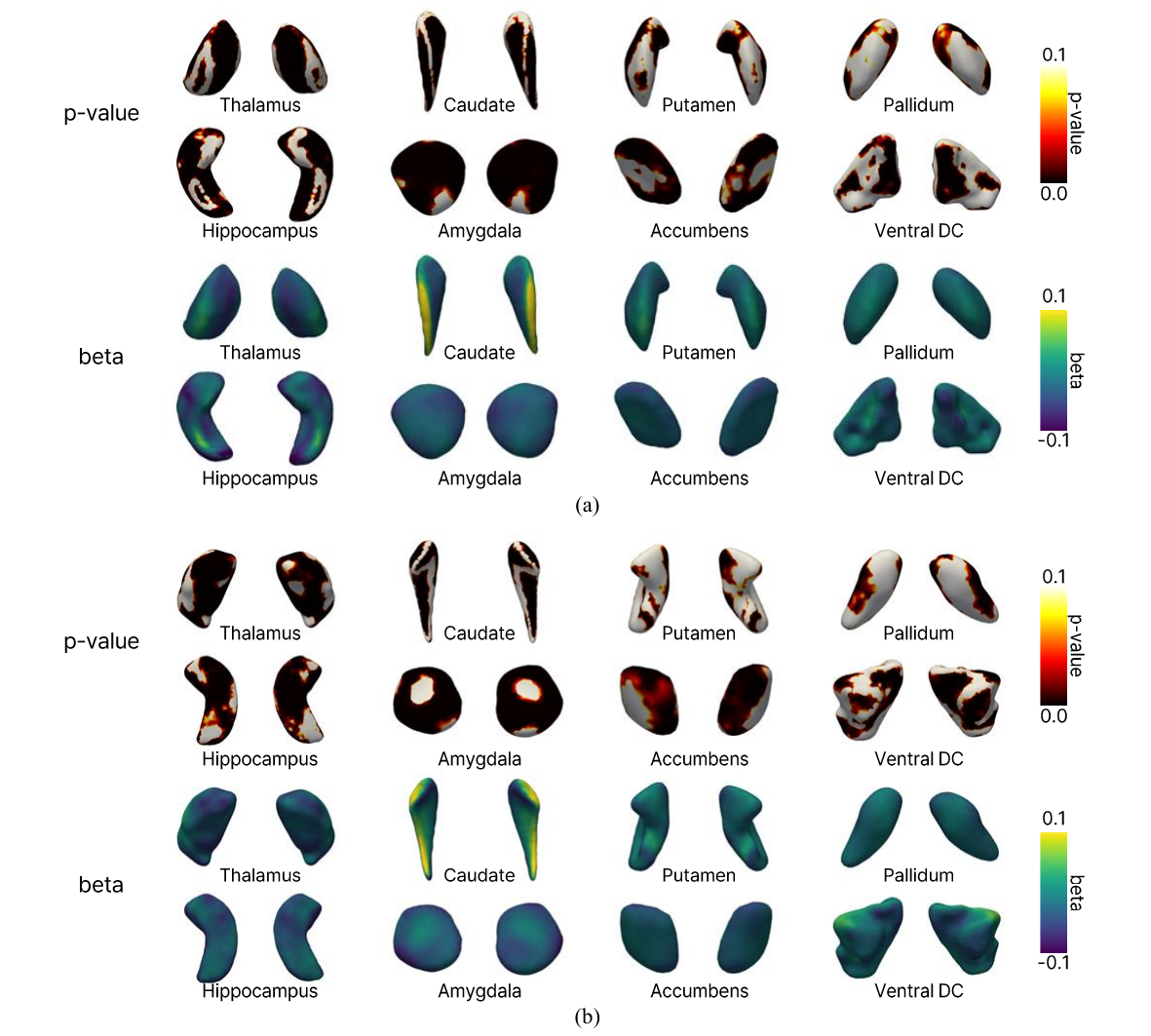}
\caption{Results from the linear mixed effect model, showing the statistical significance (above in each view, mahogany/very dark red areas) and strength (below in each view) of the association between general cognitive abilities from ages 72.5 to 82 years and shape variations in eight subcortical areas. (a) inferior-to-superior view (b) superior-to-inferior view. Positive associations (cognitive decline with outwards shape deformations) are shown as yellow areas whilst negative associations (cognitive decline, inwards shape deformations) are shown in navy (dark) blue in the shapes. Areas in green-turquoise correspond to coefficient estimates (beta) values near zero.}\label{fig:lm}
\end{figure}

\begin{figure}[t]
\centering
\includegraphics[width=1\textwidth]{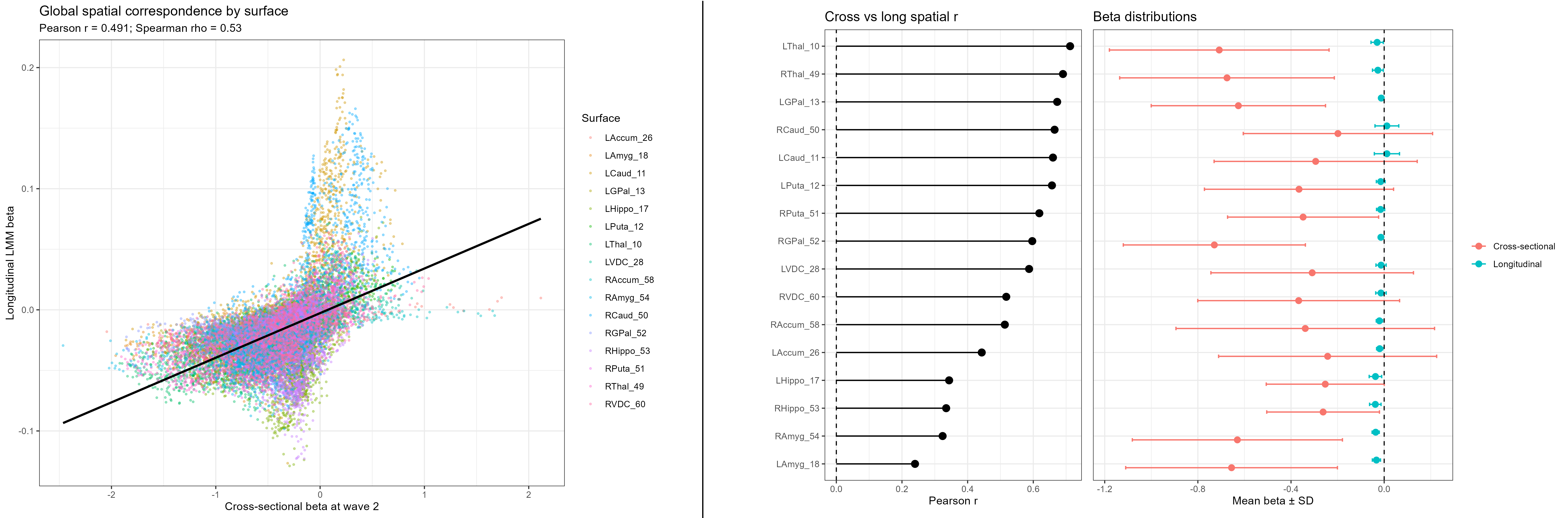}
\caption{Vertex-wise relationships between baseline cross-sectional coefficient estimates and longitudinal linear mixed model (LMM) coefficient estimates for all subcortical surfaces. Graphs show the actual model estimates (on their native scales). As such, they may partly reflect differences in model scaling in addition to differences in effect magnitudes. }\label{fig:crossvslong_correlations}
\end{figure}

\autoref{fig:lm} shows the associations between the changes in the shape morphology of the eight subcortical structures analyzed and general cognitive function (i.e., \textit{g}) across waves 2 to 5 for all subjects tested at wave 2 (i.e., baseline) accounting for childhood intelligence using a linear mixed model. As the general trend in cognitive change in this 9-year period is negative (\autoref{fig:cognition}), most statistically significant associations have a negative sign corresponding with inward shape deformations (i.e., positive slope in the vertex deformation trajectories as per \autoref{fig:vertex_in_LMM}) with respect to the average template of each structure, representing atrophy. As \autoref{tab:correlations} shows, the average coefficient estimates of these associations range between -0.038 and -0.012 (standard deviation between 0.01 and 0.02) for all structures except for the caudate, which has a positive average association ($\beta$ 0.012, standard deviation 0.05). This points at a general deformation outwards for this structure prevailing over shape contractions, as reflected also in the volumetric trajectories (\autoref{fig:volumes}).

Adding vascular risk factors and the BHI as covariates in the model did not result in noticeable changes in these associations, and neither accounting for head size, represented by the intracranial volume (ICV) in the shape modeling. Cross-sectional analyses showed variations (i.e., in the strengths as well as in statistical significance) in the associations of each mesh point deformation with general cognition at different waves. To assess the extent to which cross-sectional and longitudinal analyses captured similar spatial patterns of subcortical–cognitive associations, we quantified the vertex-wise correspondence between baseline cross-sectional coefficient estimates and longitudinal linear mixed model (LMM) coefficient estimates across all subcortical surfaces (\autoref{fig:crossvslong_correlations}, \autoref{fig:correl_individual_struc}).

Across the full subcortical system, cross-sectional and longitudinal association maps showed moderate spatial concordance (Pearson’s r=0.49; Spearman’s $\rho$=0.53; \autoref{fig:crossvslong_correlations}, \autoref{tab:correlations}), indicating that regions exhibiting stronger cross-sectional associations with cognitive function also tended to show stronger longitudinal coupling between structural and cognitive decline. However, the degree of correspondence varied substantially across structures. The strongest cross-sectional–longitudinal agreement was observed in the thalamus, caudate, putamen, and globus pallidus (Pearson’s r $\approx$ 0.60–0.71), whereas weaker concordance was evident in the hippocampus and amygdala (Pearson’s r $\approx$ 0.24–0.34).

These findings suggest that some subcortical systems exhibit relatively stable spatial organization of cognition-related structural vulnerability across cross-sectional and longitudinal analyses, whereas others may show greater divergence between baseline inter-individual differences and within-person aging-related change. Overall, size-effects of cross-sectional associations were larger than longitudinal ones with Cohen d values ranging between 0.68 and 2.32 (\autoref{tab:correlations}).

Detailed analyses of the cross-sectional and longitudinal trajectories of these associations at 3-, 6- and 9-year intervals in the individuals that contributed data are shown in the following subsections. The results from shape models after accounting for ICV are shown in \autoref{secA1} (\autoref{fig:icv_norm_thalami_cross} and  \autoref{fig:icv_norm_thalami_long} for the thalami, \autoref{fig:icv_norm_hippocampi_cross} and  \autoref{fig:icv_norm_hippocampi_long} for the hippocampi, \autoref{fig:icv_norm_caudate_cross} and  \autoref{fig:icv_norm_caudate_long} for the caudate, \autoref{fig:icv_norm_putamen_cross} and  \autoref{fig:icv_norm_putamen_long} for the putamen, \autoref{fig:icv_norm_gpallidi_cross} and \autoref{fig:icv_norm_gpallidi_long} for the globus pallidus,
\autoref{fig:icv_norm_Amygdala_cross} and \autoref{fig:icv_norm_Amygdala_long} for the amygdala, \autoref{fig:icv_norm_Accumbens_cross} and \autoref{fig:icv_norm_Accumbens_long} for the nucleus accumbens, and \autoref{fig:icv_norm_VentralDC_cross} and \autoref{fig:icv_norm_VentralDC_long} for the ventral diencephalon).


\begin{figure}[p]
\centering

\begin{subfigure}{0.75\textwidth}
    \centering
    \includegraphics[width=\linewidth]{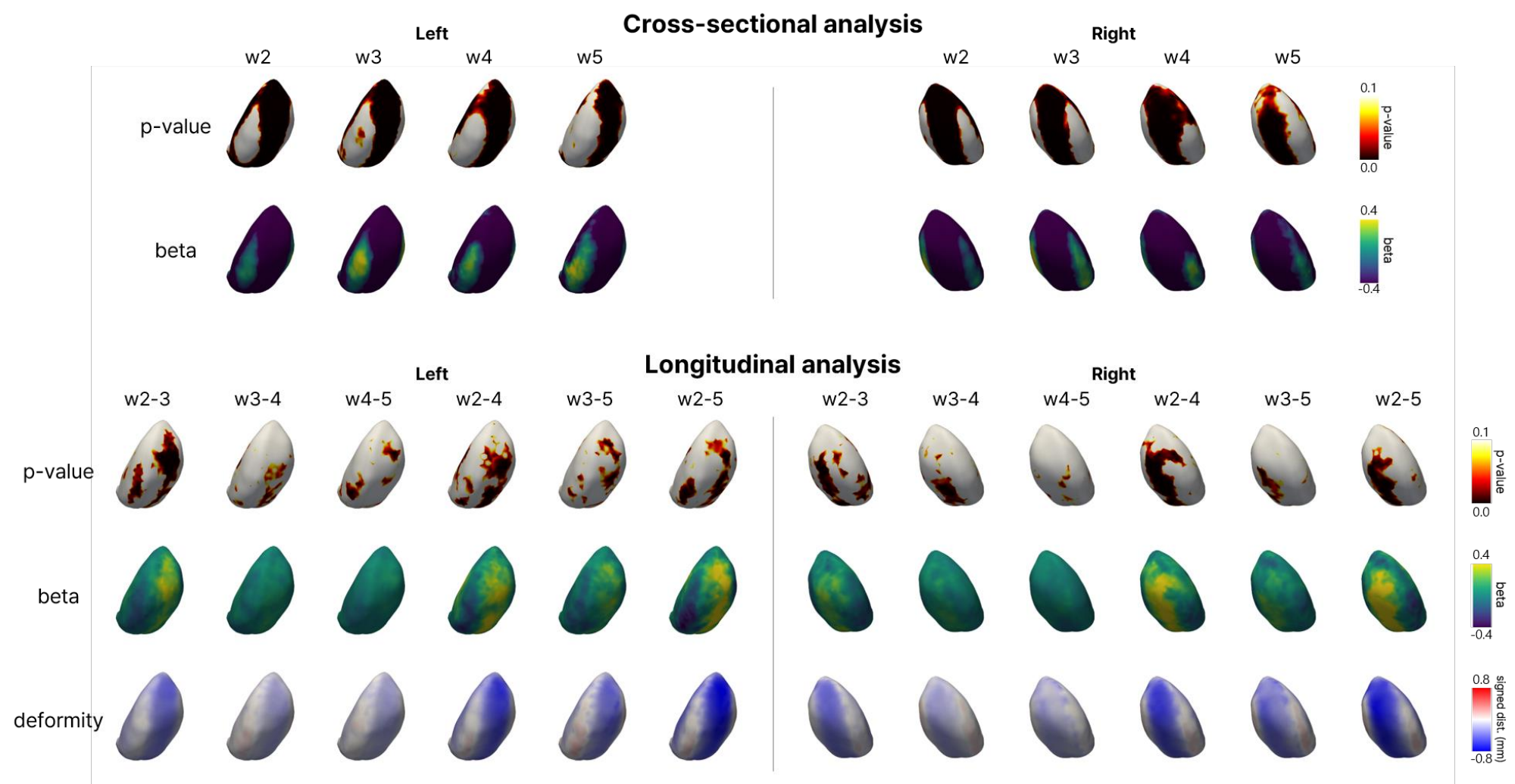}
    \caption{Thalamus}
    \label{fig:Thalamus}
\end{subfigure}

\vspace{0.8em}

\begin{subfigure}{0.75\textwidth}
    \centering
    \includegraphics[width=\linewidth]{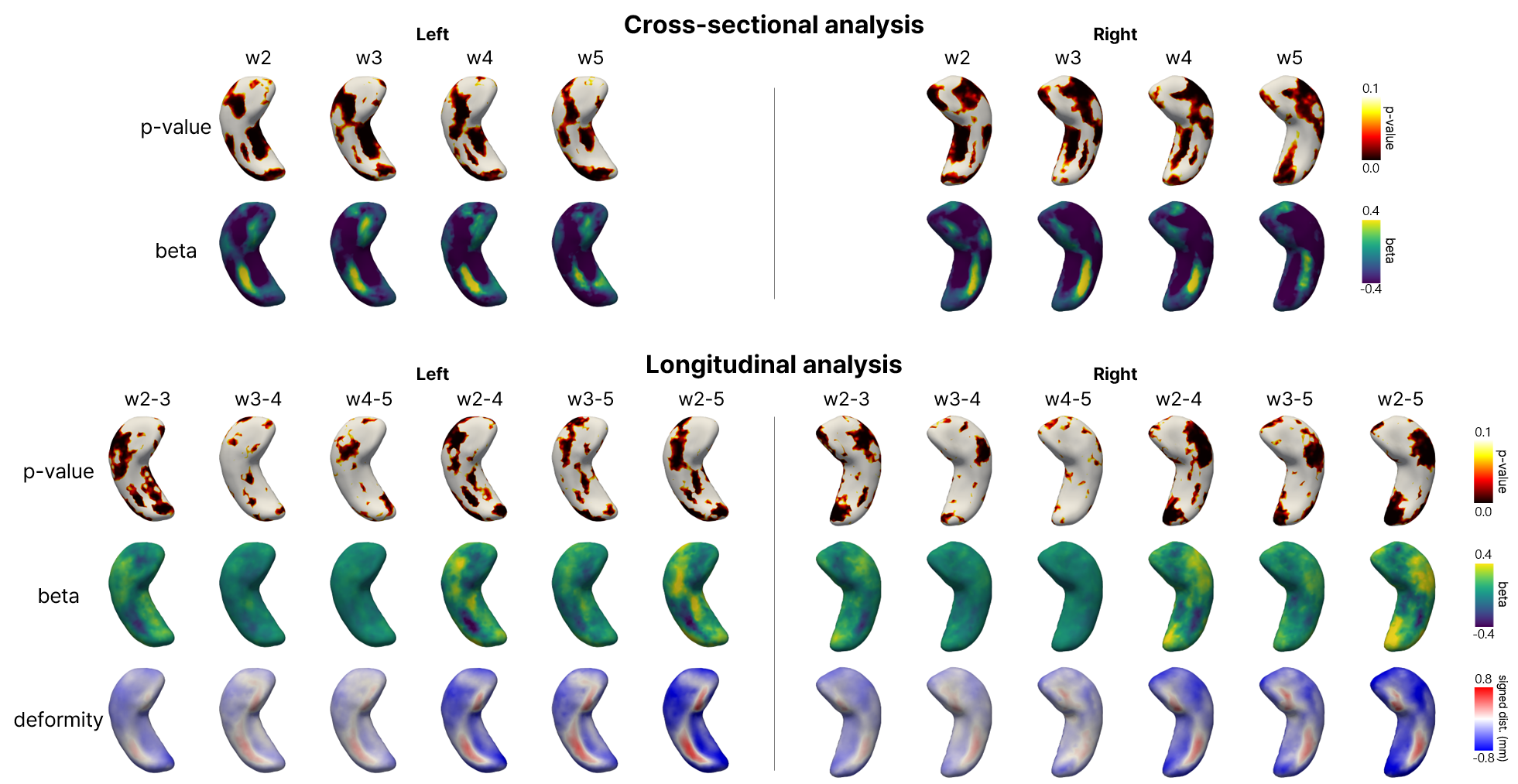}
    \caption{Hippocampus}
    \label{fig:2by2_long_hippo}
\end{subfigure}

\vspace{0.8em}

\begin{subfigure}{0.75\textwidth}
    \centering
    \includegraphics[width=\linewidth]{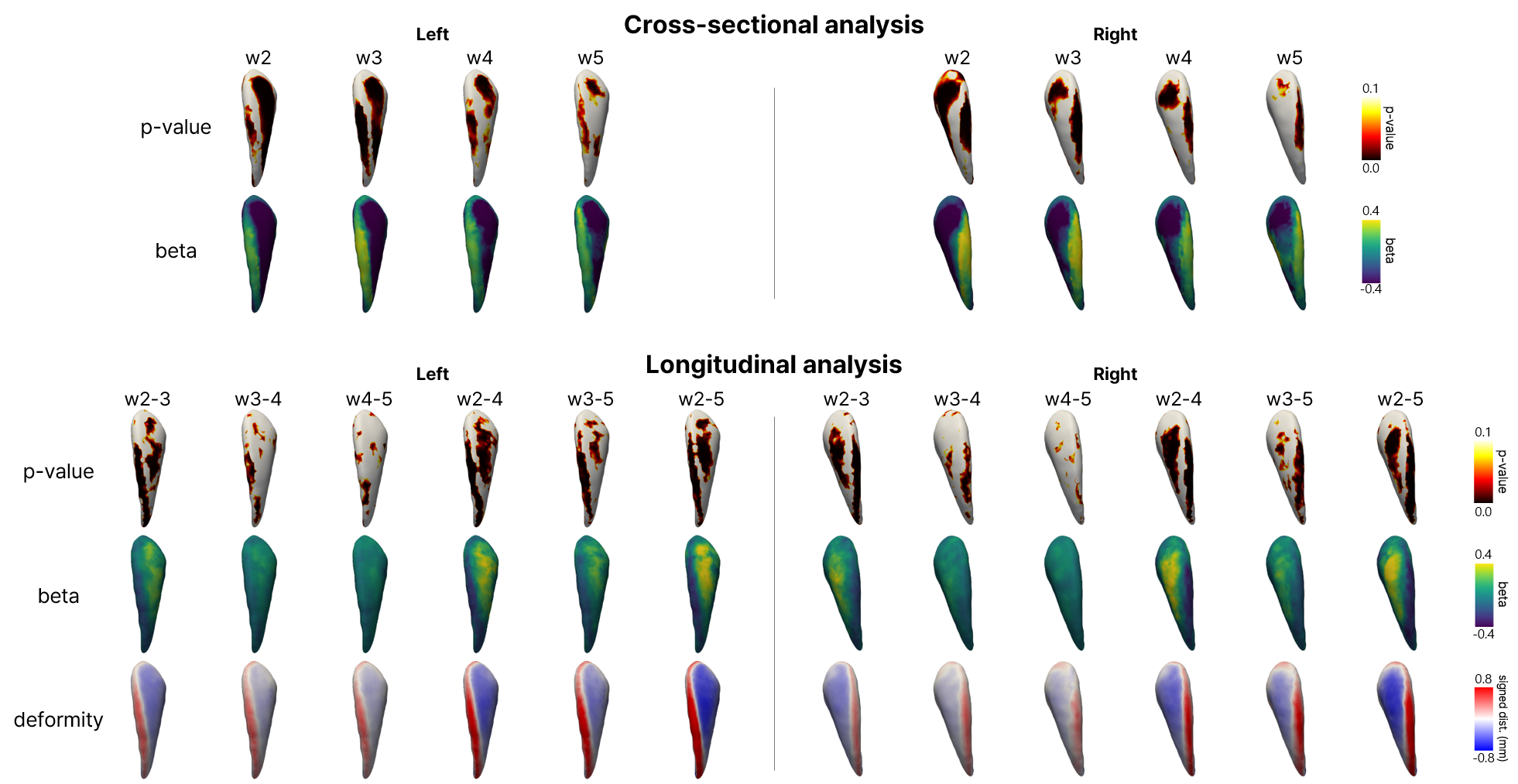}
    \caption{Caudate}
    \label{fig:Caudate}
\end{subfigure}

\caption{
Cross-sectional and longitudinal associations between subcortical shape and general cognition across aging.
The upper panels show cross-sectional associations between subcortical shape variations and cognition at waves 2, 3, 4 and 5.
The lower panels show associations between longitudinal shape changes and cognitive changes across different time intervals, accounting for childhood intelligence.
Statistical significance ($p<0.01$) is shown in dark red, positive coefficient estimates in yellow, negative coefficients in dark blue, inward deformations in blue, and enlargements in red.
}
\label{fig:subcortical_all}
\end{figure}


\begin{figure}[p]
\ContinuedFloat
\centering

\begin{subfigure}{0.75\textwidth}
    \centering
    \includegraphics[width=\linewidth]{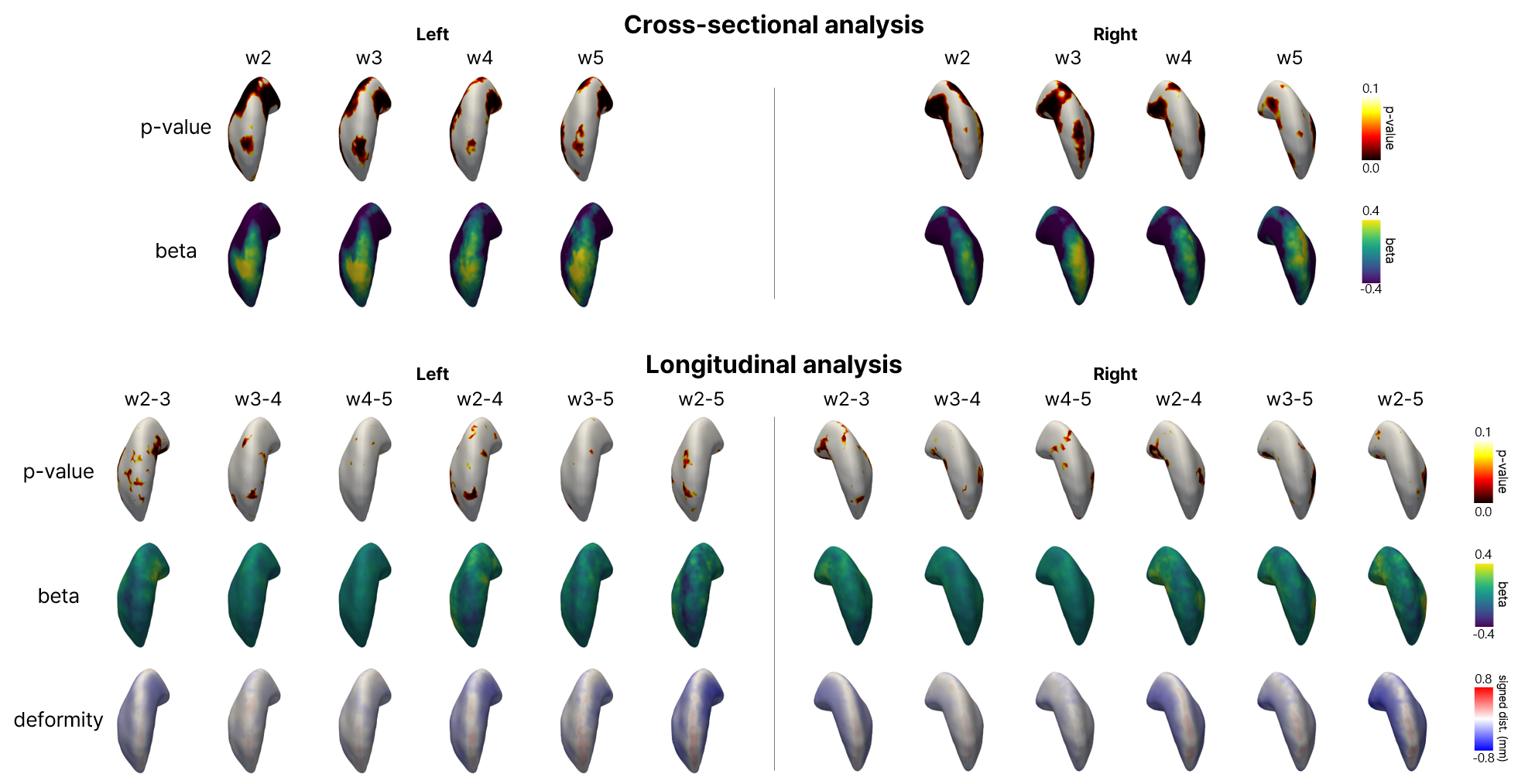}
    \caption{Putamen}
    \label{fig:Putamen}
\end{subfigure}

\vspace{0.8em}

\begin{subfigure}{0.75\textwidth}
    \centering
    \includegraphics[width=\linewidth]{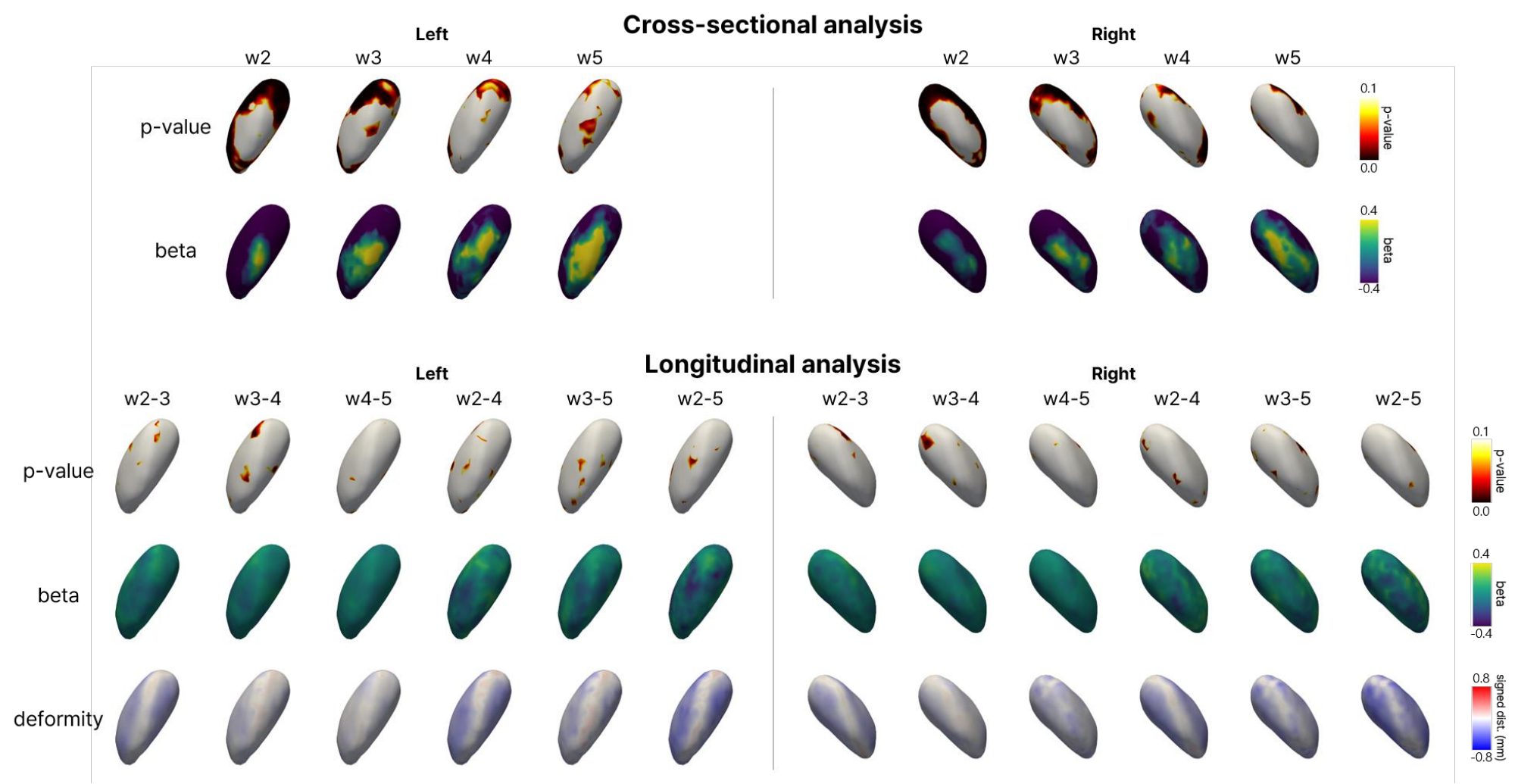}
    \caption{Globus pallidus}
    \label{fig:Gpal}
\end{subfigure}

\vspace{0.8em}

\begin{subfigure}{0.75\textwidth}
    \centering
    \includegraphics[width=\linewidth]{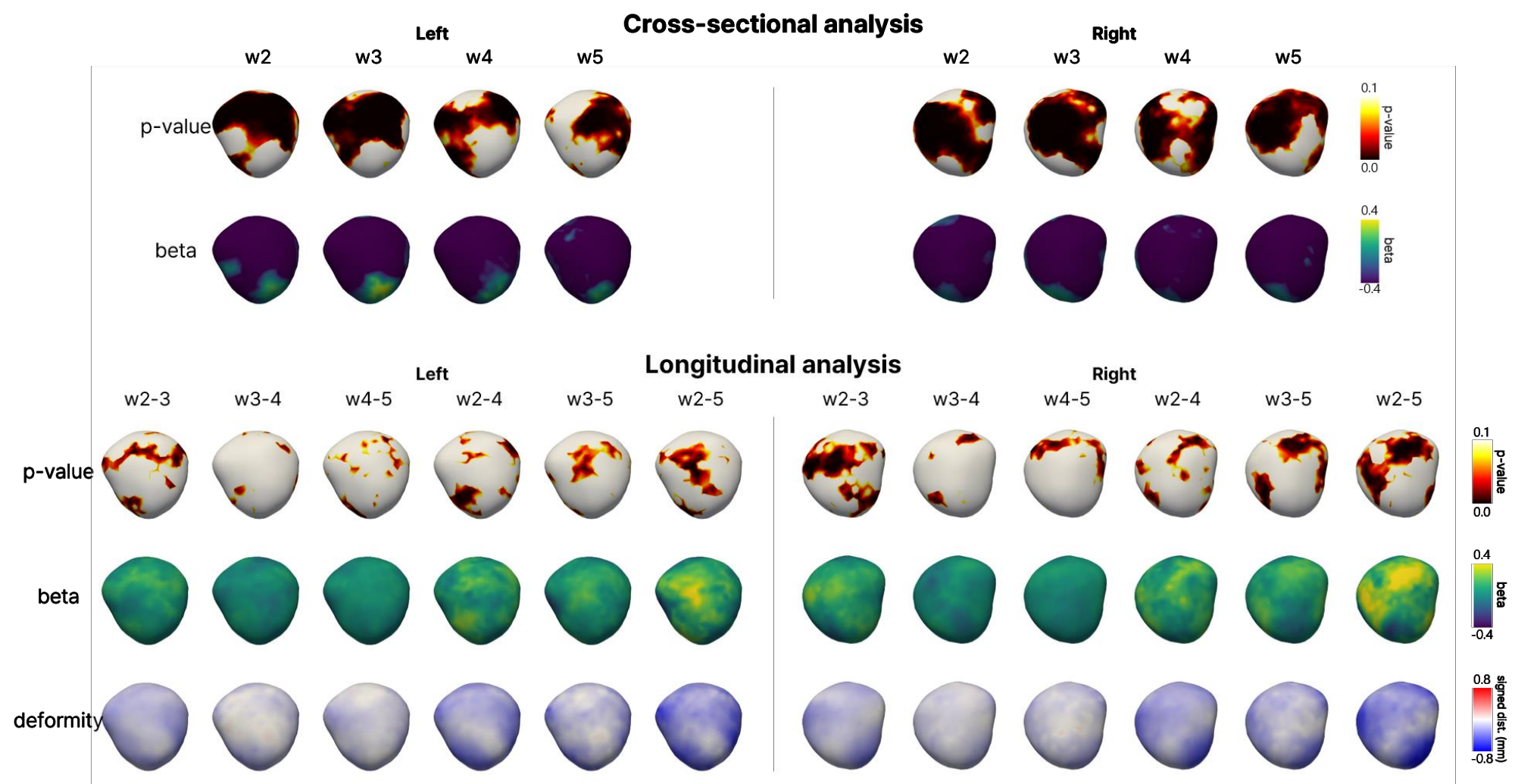}
    \caption{Amygdala}
    \label{fig:Amygdala}
\end{subfigure}

\caption[]{Figure~\ref{fig:subcortical_all} continued.}
\end{figure}


\begin{figure}[p]
\ContinuedFloat
\centering

\begin{subfigure}{0.75\textwidth}
    \centering
    \includegraphics[width=\linewidth]{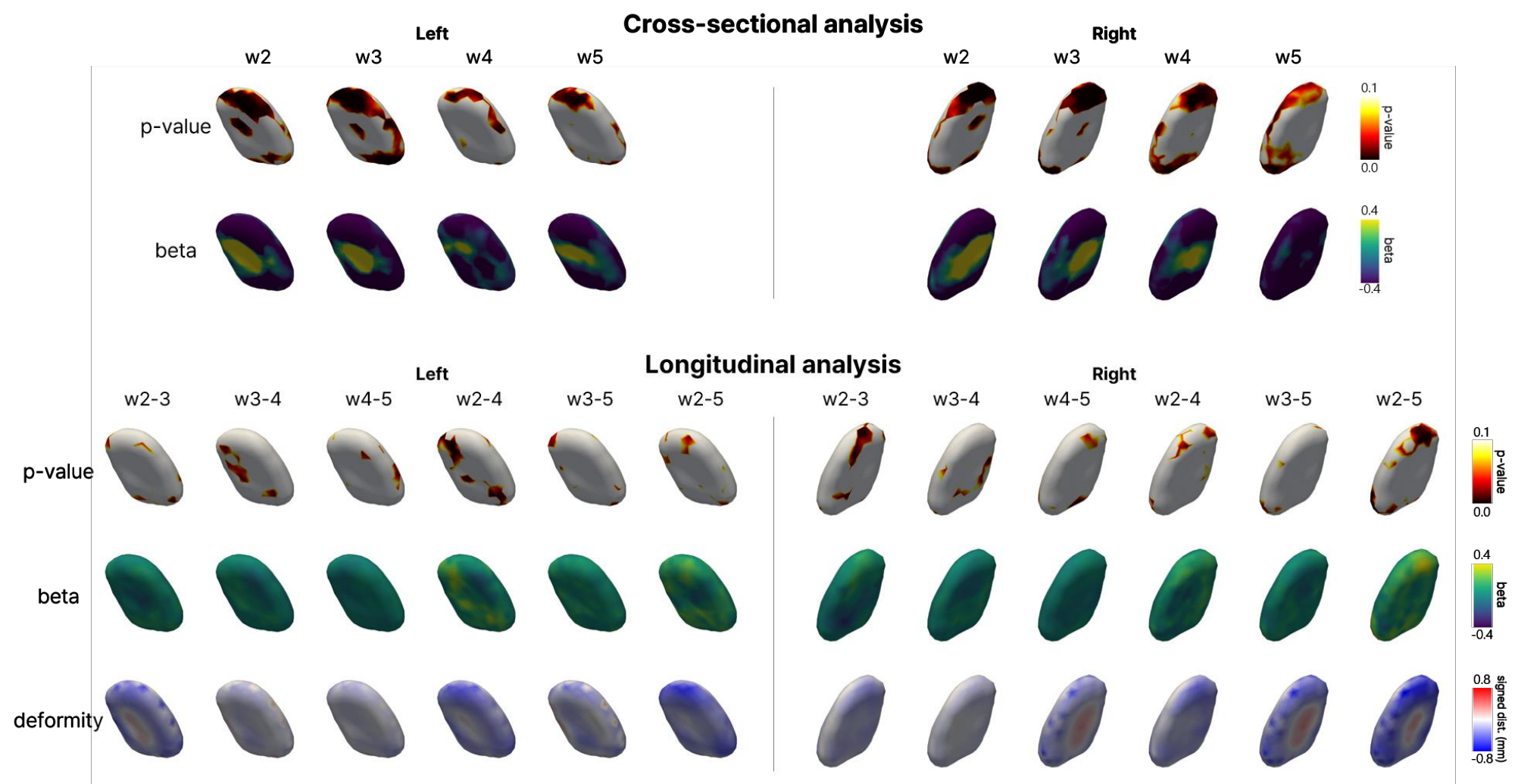}
    \caption{Accumbens}
    \label{fig:Accumbens}
\end{subfigure}

\vspace{0.8em}

\begin{subfigure}{0.75\textwidth}
    \centering
    \includegraphics[width=\linewidth]{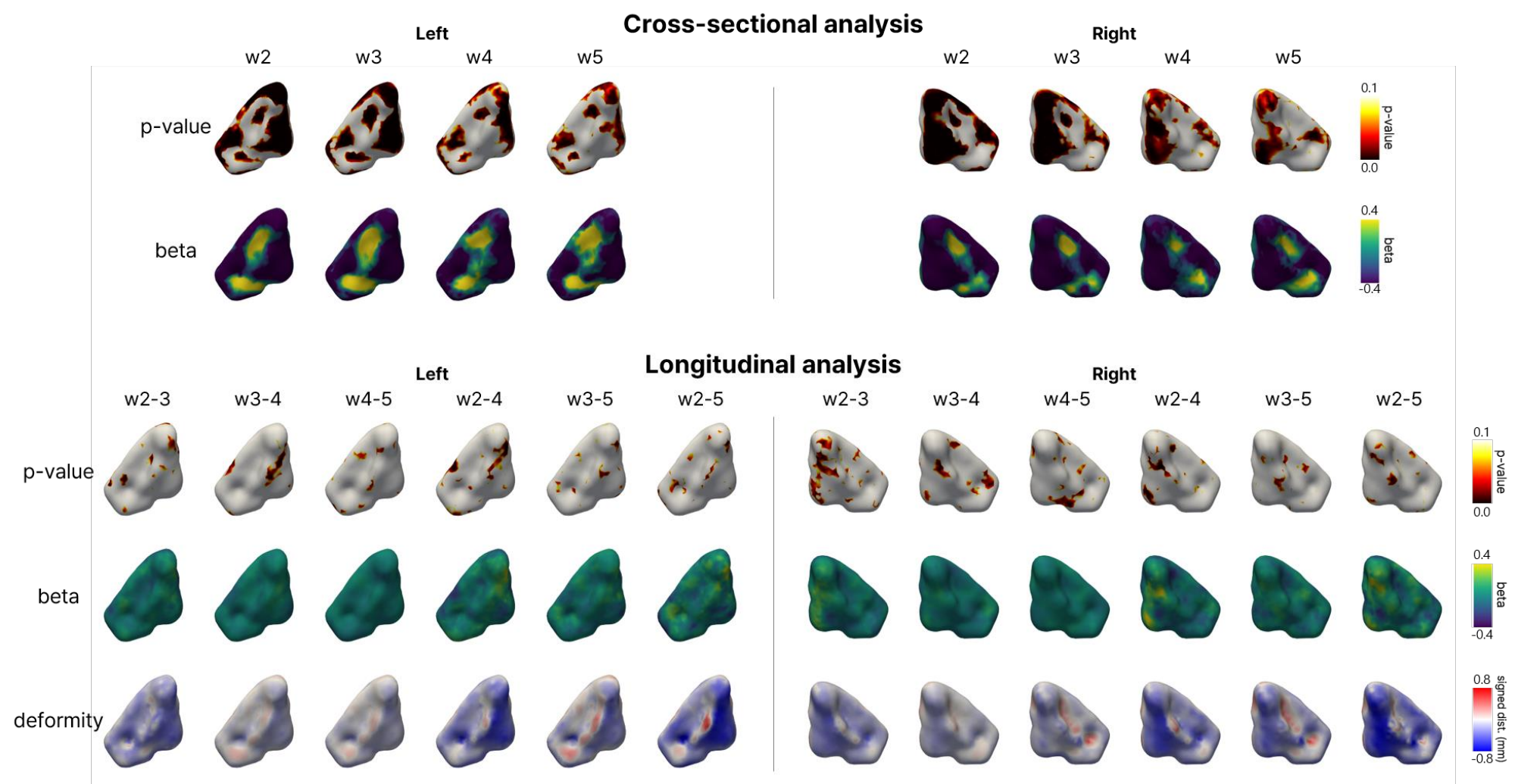}
    \caption{Ventral diencephalon}
    \label{fig:VentralDC}
\end{subfigure}

\caption[]{Figure~\ref{fig:subcortical_all} continued.}
\end{figure}

\clearpage
\subsubsection{Thalami}


The upper panel in \autoref{fig:Thalamus} shows the associations between the thalamic shape variations and general cognition in each wave. Statistical significance was achieved for both left and right thalami in the antero-medial region adjacent to the third ventricle. The right thalamus sees these areas extend more into the latero-dorsal, lateral posterior and pulvinar regions, reaching maximal coverage at wave 4 ($\sim$ 79 years old). The bottom panel in the same figure shows the associations between the change in the thalamic shapes at different intervals and the change in cognition in the same intervals for those who provided data for this calculation. As expected, the regions where these associations were statistically significant are less extensive and not uniform across the 9-year period. Shape and cognitive changes between waves 2 and 3 were the most relevant and influential in the overall 9-year associations. Although the right and left thalamus exhibited  associations at different points, they had similar patterns of changes in deformation throughout.

\subsubsection{Hippocampi}


The upper panel of \autoref{fig:2by2_long_hippo} shows the associations between the hippocampal shape variations with respect to the template and general cognition at each wave. While dorsal sections of the cornus ammonium (CA) 1 and 2 were consistently negatively associated with \textit{g}, the extent of this association in the subiculum diminished with time, especially in the right hippocampus. Adjusting the hippocampal shapes by head size (\autoref{fig:icv_norm_hippocampi_cross}) did not make any difference in the results. Associations in the CA3, CA4, and dentate gyri were statistically significant at ages $\approx$ 76 and 79 (waves 3 and 4), disappearing at $\approx$ 82 years old (wave 5). Shape variations in the dorsal head of the right hippocampus were consistently associated with cognition, but in the left hippocampus the association was restricted to a punctual area in waves 2, 3 and 4. 

\autoref{fig:2by2_long_hippo} shows the associations between the change in hippocampal shapes with respect to the average cohort shapes and cognition throughout different intervals, accounting for childhood IQ. Shape changes where these associations were statistically significant were mainly inwards with the exception of those in the dorsal CA aspect of the hippocampal tail, and varied at each interval.

\subsubsection{Caudate}


Shape changes in the time intervals of the 9-year period for the caudate can be appreciated in the bottom row of \autoref{fig:Caudate}. Unsurprisingly, shrinkage was exclusively observed in the region adjacent to the anterior horns of the lateral ventricles. Visual inspection of the binary masks and caudate shapes to explain the apparent enlargement in the region adjacent to the external capsule detected that the automatic segmentations extended the caudate shape to cover white matter hyperintensities and chronic perivascular spaces, abundant in this region. This also explains the positive slope in the volumetric change of this structure (see panel Longitudinal analysis in \autoref{fig:volumes}). Albeit in isolated points, most of the associations between \textit{g} and individual shape variations at each time point were concentrated in these two regions, suggesting rather an association with the confluence of vascular pathology adjacent to (or including) the dorsal surface of the caudate nucleus. However, shape changes across time exhibited more variation in the association with general cognition (\autoref{fig:Caudate}) .


\subsubsection{Putamen}

For the sample of individuals who provided data for the analyses, individual putaminal shape variations at the different time points showed consistent patterns of associations with general cognition, significant in the anterior part adjacent to the internal capsule and in the dorsal inferior surface adjacent to the external capsule, more prominent in waves 2 and 3. However, longitudinal shape changes between time-points in the 9-year period in left and right putamen,  were not associated with cognitive changes in general (\autoref{fig:Putamen}), with the exception of some small clusters in the dorsal inferior boundary possibly rather associated with chronic lesions of suspected vascular origin (i.e., mainly lacunes and strings of perivascular spaces) in the inferior portion of the external capsule. 

\subsubsection{Globus Pallidus}


For the globus pallidus, cross-sectional associations between shape variations at each time point were at their border with the internal capsule and the putaminal head at ages $\approx$ 72.5 and 76 (waves 2 and 3), and in small clusters scattered throughout at ages $\approx$ 79 and 82 (waves 4 and 5, cross-sectional analysis panel in \autoref{fig:Gpal}). However, similarly to the putamen, longitudinal shape changes in the globus pallidus, in general, were not associated with cognitive changes in the time intervals comprised in the 9-year period for subjects who provided data (bottom panel Longitudinal analysis in \autoref{fig:Gpal}).  

\subsubsection{Amygdala}


\autoref{fig:Amygdala} shows the associations between shape variations (upper panel) and changes (bottom panel) of the left and right amygdala and general cognition. Interestingly, structural changes in this structure were limited to shrinkage (i.e., outwards shape deformations or regional shape enlargements were not observed), and association patterns markedly differed between left and right. Cross-sectionally, associations with cognition in the left amygdala were in its central, medial and cortical regions, while in the right they also extended through the lateral and basolateral regions. The associations with the longitudinal changes in the time intervals within the 9-year period (see w2 to 5 in \autoref{fig:Amygdala} ) were restricted to the basolateral region in the left while also including the other regions in the right, preceded by associations in scattered clusters that differed in position and extent between both hemispheres.

\subsubsection{Accumbens}


The nucleus accumbens shows association patterns consistent across the waves and limited to a small region adjacent to the caudate (\autoref{fig:Accumbens}). Longitudinal vertex-wise changes across the different time intervals were only associated with general cognitive changes in very small points, with small effect size.

\subsubsection{Ventral DC}


\autoref{fig:VentralDC} shows the association patterns for the ventral diencephalon. As the accumbens, globus pallidus and putamen, the short-term morphological changes in this structure were associated with general cognitive change only in small scattered regions. However, cross-sectionally, shape variations with respect to the reference template were associated with general cognition in large clusters, mainly in the hypothalamic fissure and on the surface area limiting the third ventricle. 


\subsection{Sensitivity analysis}

To investigate whether differences in the shape modeling process would influence the results, and if so in which way, we conducted a sensitivity analysis with the hippocampal shape modeled using a template generated previously \cite{park2025ai} using hippocampal binary masks from 654 LBC1936 participants (at scan 1, wave 2), available from \url{https://doi.org/10.7488/ds/7874}, but remeshed with 2588 vertices. Hence, we repeated the analyses described above using a template mesh with a higher number of vertices (i.e., 2588 for left and right hippocampus) than the mesh used for the main analyses (i.e., 1490 vertices for left hippocampus and 1492 vertices for right hippocampus, as per \autoref{tab:number_of_vertices}).

\begin{figure}[H]
\centering
\includegraphics[width=1\textwidth]
{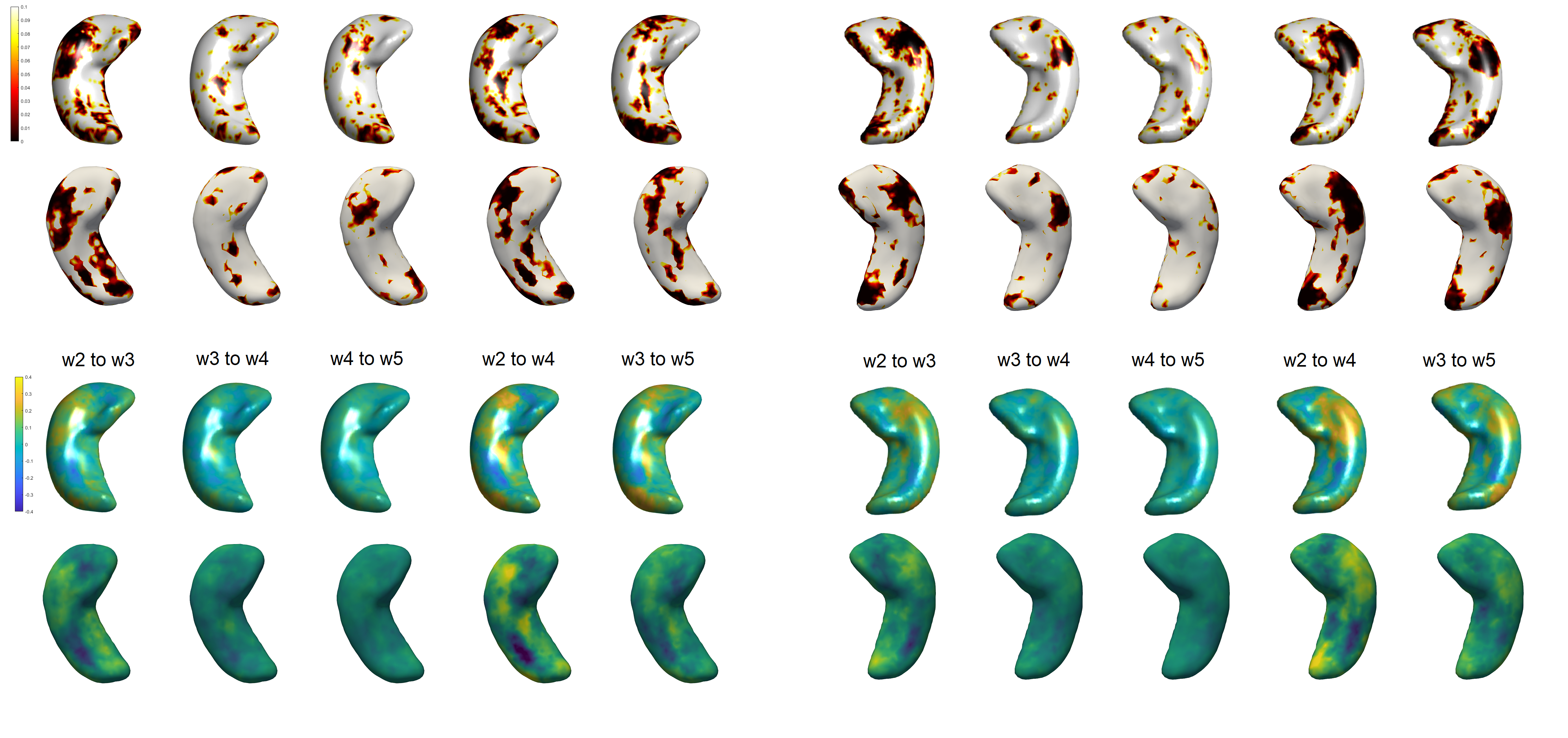}
\caption{Sensitivity analysis. Associations of \textit{g} and hippocampal shape deformations in different time intervals, from using two different template meshes: 2588 vertices (row above) and 1490 vertices (row below). The color scales (for the p and B values) are the same for both }\label{fig:sensitivity_long_comparison}
\end{figure}

\autoref{fig:sensitivity_long_comparison} allows us to visually compare the associations of \textit{g} and hippocampal shape deformations (accounting for age 11 IQ) in different time intervals obtained using the template mesh with 2588 vertices (upper rows), against those presented in \autoref{fig:2by2_long_hippo}, which uses a template mesh with nearly half the number of vertices (rows below). Despite noticeable differences in the template meshes, the anatomical regions where the associations are statistically significant are consistent regardless of the number of vertices in the template mesh. There is also consistency in the strength and direction of the association throughout the 3D shapes regardless of the template used.

\begin{figure}[H]
\centering
\includegraphics[width=0.6\textwidth]
{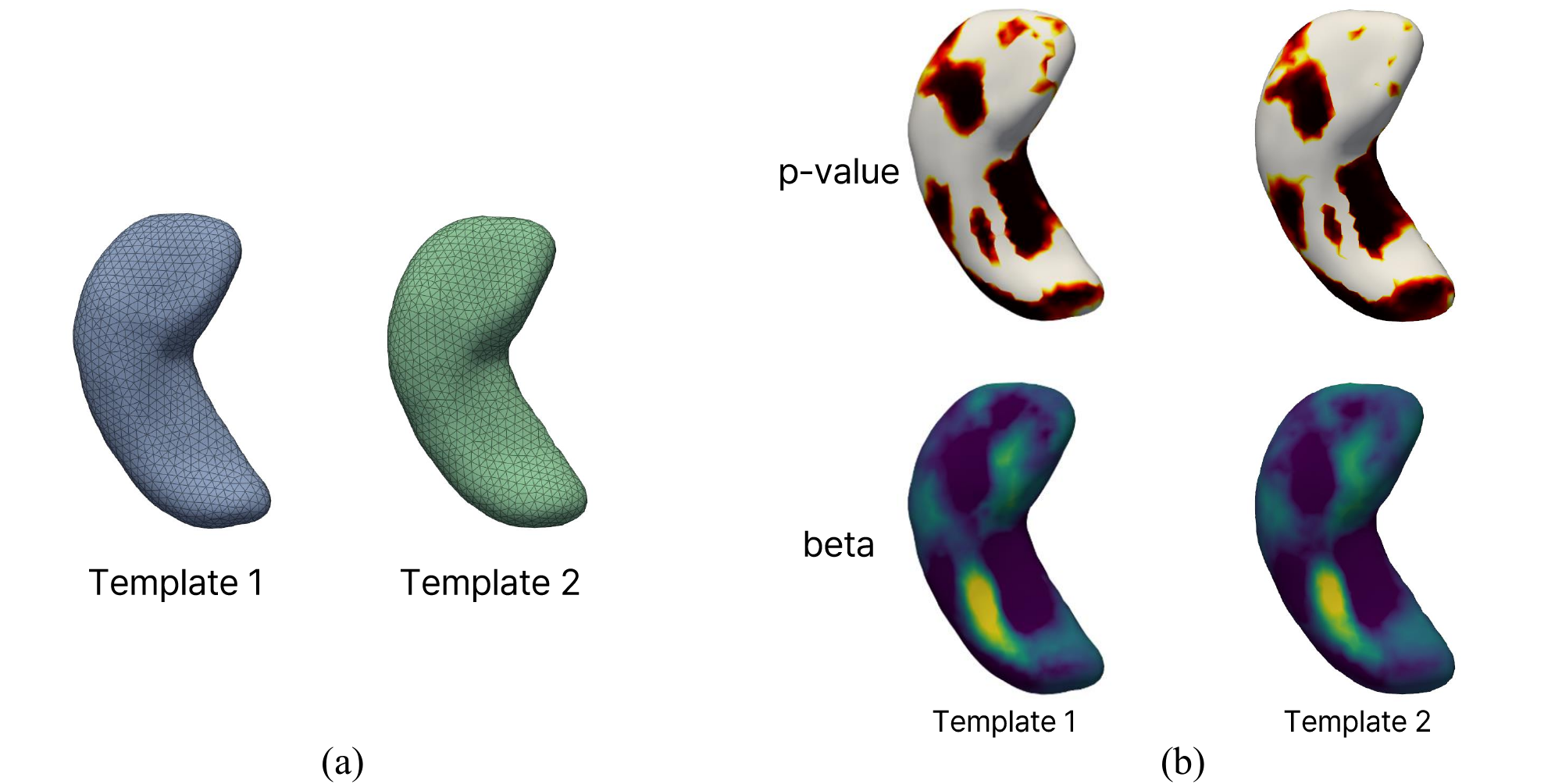}
\caption{Sensitivity analysis of template selection. (a) Two template meshes constructed from two independently and randomly sampled sets of 100 subjects. (b) Corresponding statistical maps (p-values and beta coefficients) obtained using the two different template meshes at Wave 2.}\label{fig:sens_temp}
\end{figure}

We further assessed the sensitivity of our framework to the choice of template mesh in \autoref{subsec:template_construction}, as illustrated in \autoref{fig:sens_temp}. To generate each template, 100 subjects were randomly sampled, and their shapes were averaged to construct the reference mesh. As shown in (a), although the templates were derived from entirely different subject sets, their overall geometric features are highly similar.

Importantly, despite minor differences between template meshes, the optimization procedure based on the algorithm proposed in \cite{park2025ai} enables precise reconstruction of individual subject-specific shapes. Consequently, the resulting clinical association analyses remain nearly identical across template choices, as demonstrated in (b).

To further corroborate consistency in the results regarding the regions where shape variations were (statistically significantly) associated with cognition at each time point, we also compared the shape variation patterns from two groups: those that had \textit{g} values above the mean + standard deviation (i.e., higher \textit{g}) \textit{versus} those that had \textit{g} values below the mean - standard deviation (i.e., lower \textit{g}). We used the Wilcoxon rank sum test (Mann-Whitney U) for these group comparisons.

\begin{figure}[H]
\centering
\includegraphics[width=1\textwidth]
{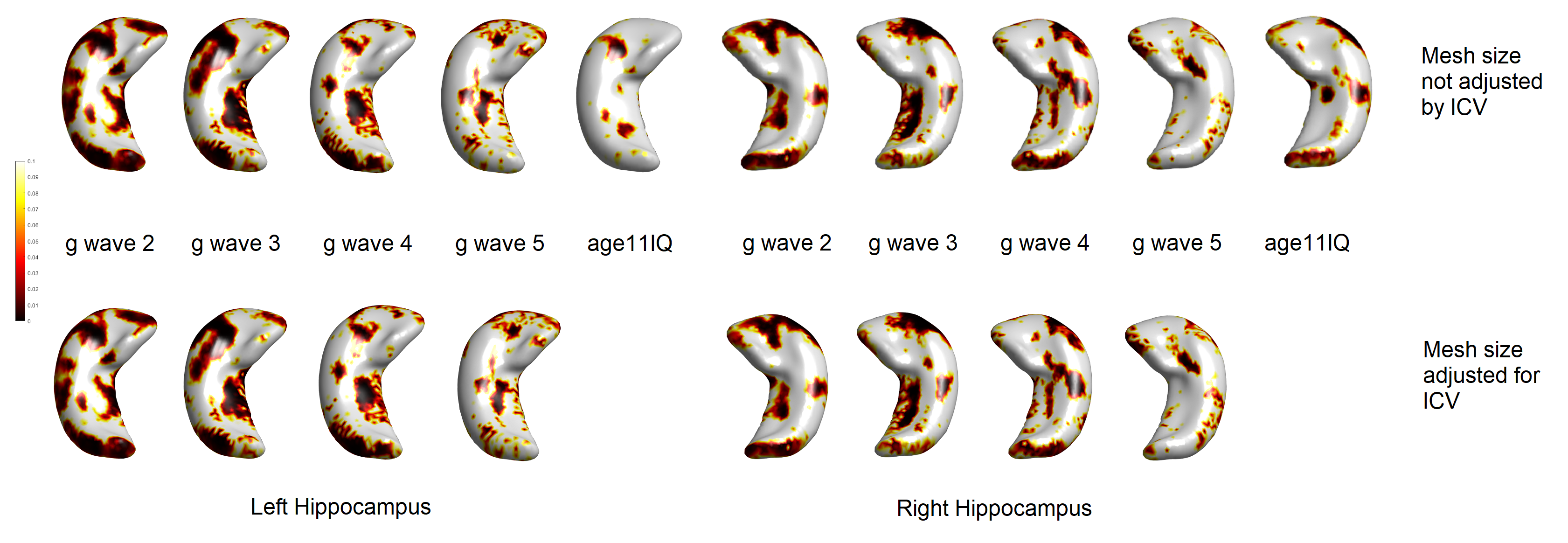}
\caption{Sensitivity analysis. Results from comparing the hippocampal shape deformations at each time point in individuals with higher levels of \textit{g} and those with lower values. Use a template of 2588 vertices.}\label{fig:sensitivity_cross}
\end{figure}

\autoref{fig:sensitivity_cross} shows the results from using the Mann-Whitney U test to compare the shape deformations of individuals with higher values of \textit{g} against those with lower values of \textit{g} at each wave, and the individuals with high IQ at age 11 against those with low IQ at that age. Shape differences in the subiculum and CA1 diminish  with time, and in the CA2 and at the tip of the hippocampal tail disappear at 82 years (wave 5). Interestingly, group differences at 82 years old are similar to those at age 11. Adjusting the mesh size for head size did not visibly change the results. This interesting result points at the more determinant link between brain morphology and childhood fluid intelligence in the ninth decade of life compared to the previous decade.

\autoref{fig:sensitivity_long} and \autoref{fig:sensitivity_long_ICV} show the effect of adding different covariates in the statistical models that explored the longitudinal associations between \textit{g} and hippocampal shape changes in different time-intervals. Adding vascular risk factors, BHI (or BPF), and sex did not visibly change the strength, direction, or significance in any of the associations, and neither adjusting the 3D shapes by head size prior to computing the shape deformatioons with time. Differences were limited to isolated points, encircled in the top row of \autoref{fig:sensitivity_long_ICV}, and with very small effect sizes, likely to disappear after correcting for multiple comparisons.

\section{Discussion}\label{sec:discussion}

As hypothesized, individual subcortical shape and volume changes are heterogeneous, with varied atrophy patterns across the 8th decade of life, i.e., some regions at some point looking enlarged while others shrink. Overall, shrinkage prevails in the longitudinal deformation patterns, but hippocampus and caudate also showed small areas of outward shape projections compared with the relative vertex location of the shape boundaries of the previous scan. These shape morphology changes were not uniform throughout the 9-year period analyzed. The structural morphological changes in all structures were associated with changes in general cognition, which confirms that regardless of the specific cognitive functions in which each of them intervenes, all are part of the neuronal circuits that underpin general cognitive abilities.

One would probably not have expected the cross-sectional associations between vertex-wise deformations and general cognition at different waves to be much different, and neither the change - change associations between specific time points to be much different to the change - change associations when all time points are considered.  However, our analyses show that although considering wide subregions within these structures these expectations held, the size and location of the center of these clusters where associations were statistically significant varied throughout the 9-year period. When multiple measurements were used to estimate the associations between the changes throughout the entire time-frame (i.e., in the linear mixed models), the results lacked insight into the fluctuations at specific spatial locations and in specific intervals. 

A notable finding was the differing degree of spatial correspondence between cross-sectional and longitudinal cognition-association maps across subcortical structures. While thalamic and striatal surfaces showed relatively strong agreement between cross-sectional and longitudinal spatial patterns, substantially weaker correspondence was observed in the hippocampus and amygdala. This heterogeneity is potentially informative. The thalamus has repeatedly been implicated in general cognitive functioning and cognitive aging, including in prior work from the present cohort and others \cite{cox2016ageing}, where it has emerged as a particularly robust subcortical correlate of cognitive ability. So has the globus pallidus \cite{hernandez2020striatum}. By contrast, hippocampal and amygdala associations may be more sensitive to distinctions between cross-sectional inter-individual variation and within-person longitudinal change, potentially reflecting the greater temporal and pathological specificity often attributed to medial temporal lobe structures in aging and neurodegeneration research. Together, these findings suggest that the spatial organization of cognition-related subcortical associations is not uniform across structures, and quantify how far cross-sectional and longitudinal analyses may capture partially overlapping but non-identical aspects of brain–cognition relationships in later life.

The shape modeling analysis methodology is another strength in the study. Two previous works have compared subcortical shape deformation trajectories between controls and disease groups. One study (\cite{muralidharan2014diffeomorphic}) estimated diffeomorphic shape trajectories from segmented shapes across time to produce a continuous and temporally consistent sequence of shapes for exploring differences in shape trajectories between Huntington's disease patients and a group of healthy individuals. The other study \cite{laansma2024worldwide} compared the 3D shapes of the subcortical structures of Parkinson's disease patients at different disease stages against healthy controls using two metrics: the distance of the vertices to a medial curve which for a cylindrical-like shape is its center axis (referred in the publication as "thickness"), and the logarithm transform of the ratio of the triangular area relative to the triangular area in the template at corresponding vertices referred to as "contraction or expansion". These modeling strategies have not been applied to analyze continuous changes in an age-homogeneous cognitively normal sample, and, therefore, their susceptibility to capture small variations and temporal changes while being robust to imprecisions in segmentation and variations in image resolution and template design choices is not known.

Our modeling method, as \cite{muralidharan2014diffeomorphic}, considers segmentation inconsistencies \cite{park2025ai} and, without losing sensitivity, proved robust against them. In \cite{park2025ai} neither the density or distribution of the vertices nor the choice of template mesh influenced the associations between shape deformations at each vertex and the different cognitive domains evaluated. Statistical comparison of medians and standard errors of the associations of the deformation values for each vertex and general cognition throughout the multiple experiments in \citep{park2025ai} revealed that only the binary mask segmentation method (i.e., manual segmentation vs. SynthSeg automatic segmentation) influenced the results in surface regions that had borderline statistical significance.

We used binary masks from the subcortical structures of 100 brain scans from wave 2, randomly selected, to construct the templates. It can be argued that while random sampling aims to reduce bias, it does not guarantee the sample perfectly represents the population, especially without a previous analysis of follow-up participation from the subsample randomly chosen. Also, it can be hypothesized that there would be variability in the overall results using data from a randomly selected subsample, as different random samples may produce different outcomes. However, the results of our sensitivity analysis testing the possibility of our selection choice affecting the final shape, yielded negligible variations in the number and relative positions of the vertices. Further analyzes using different templates with considerable differences in the number of vertices did not affect the pattern, direction, and statistical significance of the associations between cognition and subcortical shape changes.

To make the template for each target structure, the first step was to rigidly align the binary masks of these structures from 99 subjects to the corresponding binary masks from the first subject of those 100 randomly selected. It is common practice to generate templates in the MNI-152 space, referred as standard space. However, rigidly registering all T1-weighted brains (from where the masks were derived) to the MNI-152 brain template does not warrant that their subcortical structures are aligned. We are aligning one by one, separately, each subcortical structure to construct their templates (i.e., one template per structure). Alternatively, performing non-rigid registrations that would warp simultaneously all subcortical structures to a standard brain template would erase all individual variations needed to be captured for generating the cohort templates of each subcortical structure. Moreover, it is worth noting that  the position of the 3D object (i.e., template mesh) in the image space is irrelevant for the shape analyses performed, as the relative position between the subcortical structures' shapes is not analyzed.

We use the vertex-wise shape variations (i.e., deformations) with respect to the average template and compensate for missed values using mixed-effect models. These models estimate dropout trajectories by utilizing all available longitudinal data to construct both population-level and individual-level trends assuming that data is missing at random. All participants who attend cognitive testing are invited to also undertake an MRI brain scan, even if they may have memory concerns or even if they have been diagnosed with a form of dementia. But participation is on a voluntary basis, and a proportion of participants have chosen not to have an MRI at each wave. Although this decision is not necessarily due to dementia diagnosis or worsening in cognitive status, missing values analysis showed that, in general, participants who declined to participate in subsequent waves had worse cerebrovascular disease and worse general cognition in the waves they contributed to than those who contributed to the subsequent analyses, hinting at these patterns of associations found being representative of healthy aging rather than of the general elderly in the 8th decade of life.

Despite these limitations, we have shown, for first time, how subcortical brain structures experience dynamic and non-uniform morphological changes associated with fluctuations in general cognitive performance across the 8th and beginnings of the 9th decade of life in community-dwelling individuals. We also show that not all subcortical structures "age" at the same speed and neither these "aging" patterns are consistently associated with the individuals' general cognition. The hippocampus and the ventral DC experienced morphological deformations that differed in left and right hemispheres, while the thalami and globus pallidi shapes experienced a more uniform volume contraction in the 9-year period studied, nearly symmetrical throughout the different 3-year periods in which we subdivided the decade. We make available all the data, analyses, shape templates, sure that the wealth of information generated in this study will be useful in studies of cognitive aging and cognitive epidemiology.

\section{Methods}\label{sec:method}

\subsection{Subjects}

The Lothian Birth Cohort 1936 (LBC1936) Study, comprises cognitively normal community-dwelling individuals of single ancestry, who had undergone extensive cognitive testing and brain MRI multiple times. Participants were all born in 1936 in Edinburgh and the Lothians central-east region of Scotland at recruitment (2004). They took the Moray House Test No. 12 - a standardized intelligence test - as part of the Scottish Mental Survey of 1947, at age 11 years. At approximately 70 years of age, they formally consented to participate in the LBC1936 Study, a longitudinal study that primarily aims at exploring how childhood intelligence relates to cognition and health in older years. As part of the LBC1936 Study, participants have been undertaking the same Moray House test and an additional battery of cognitive tests every three years. From the second wave of tests at the average age of 72.5 years onward, participants also have undergone brain MRI. This study uses data from this (i.e., wave 2) and the following three assessments at mean ages of 75.5, 78.5, and 81.5 years. The study protocols were approved by the Multicentre Research Ethics Committee for Scotland (MREC/01/0/56), the Lothian Research Ethics Committee (LREC/2003/2/29), and the Scotland A Research Ethics Committee (07/MRE00/58).

\subsection{MRI acquisition}

Brain MRI was performed at the Signa GE 1.5T research scanner at the Western General Hospital in Edinburgh following the same acquisition protocol,  described in \cite{wardlaw2011brain}. For this study we used imaging data derived from the T1-weighted MRI, a 3D inversion recovery (IR) preparation sequence combined with a fast spoiled gradient recalled (FSPGR) echo acquisition with TR/TE/TI of 10/4/500 ms, matrix of 192 x 192 zero padded to 256 x 256, and voxel size of 1 x 1 x 1.3 mm$^{3}$.

\subsection{Shape Modeling}

\subsubsection{Image Processing}

We obtained binary masks for all subcortical structures fully automatically using SynthSeg \cite{billot2023synthseg}, a deep learning method that received as input the T1-weighted volumes to output a composite segmentation mask. From this composite mask, we extracted separate binary masks for each subcortical structure, which were visually checked for accuracy. 

\subsubsection{Template mesh modeling}\label{subsec:template_construction}

\begin{figure}[h]
\centering
\includegraphics[width=1\textwidth]{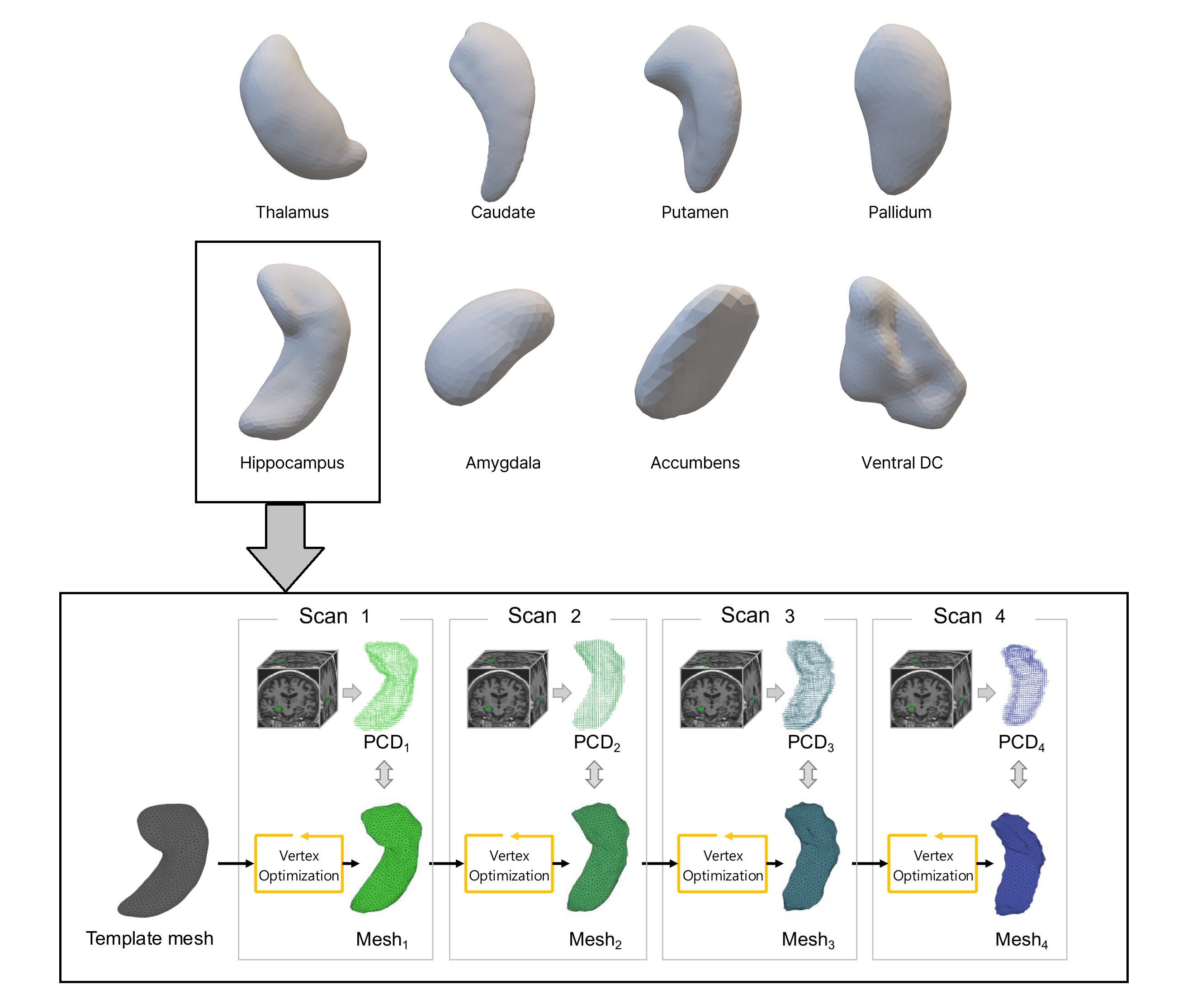}
\caption{Template meshes of the eight subcortical structures and schematic illustration of the longitudinal shape reconstruction process from the hippocampal template mesh as example. PCD: point cloud distribution.}\label{fig:temp_mesh}
\end{figure}

From Wave 2 of the LBC1936 dataset, we randomly selected 100 subjects and extracted binary masks of their subcortical brain structures. This sample size was chosen to ensure a stable and representative group-average shape while maintaining computational efficiency during the construction of the mesh. To assess the robustness of the template construction with respect to subject selection policy, we performed a sensitivity analysis using independently sampled subject sets (see \autoref{fig:sens_temp}), which confirmed minimal variation in downstream statistical results. 

For each target structure, masks from 99 subjects were rigidly aligned to the first subject (from those randomly selected) using the iterative closest point (ICP) algorithm~\cite{besl1992method}. Voxels occupied by more than 30\% of subjects were retained to define template occupancy, ensuring structural connectivity of the resulting mesh. This threshold of 30\% was chosen based on the proportion that, empirically, would be outside of one standard distribution above and below the mean of a normal probability density function. Based on this occupancy map, an initial template mesh was generated using the marching cubes algorithm, followed by surface smoothing with the Taubin method~\cite{taubin1995curve}. Finally, the template was refined through remeshing to achieve an average edge length of approximately 1.5 mm, as illustrated in \autoref{fig:temp_mesh}. \autoref{tab:number_of_vertices} shows the number of vertices for each template mesh.

\subsubsection{Individual longitudinal shape reconstruction}

The individual shape modeling of each subcortical structure was performed using the AI-based deformable mesh optimization framework~\cite{park2025ai} and longitudinal shape modeling method~\cite{parkbrainode}. As illustrated in \autoref{fig:temp_mesh}, for each structure, the first time-point mesh was reconstructed by deforming the structure's template mesh to fit each individual first time-point MRI scan (i.e., at wave 2). Prior to deformation, the corresponding point cloud \( \mathrm{PCD}_1 \) was rigidly registered to the template mesh using, as previously mentioned, the ICP algorithm~\cite{besl1992method}. Subsequently, meshes at the following time-points were reconstructed sequentially by deforming the reconstructed mesh at the previous time-point. Specifically, each target point cloud \( \mathrm{PCD}_N \) was first rigidly aligned to \( \mathrm{PCD}_{N-1} \) using ICP before deformation.

The AI-based optimization framework iteratively updates vertex positions over 5000 iterations using an MLP-based PointNet architecture~\cite{qi2017pointnet}. The optimization is guided by a distance loss that aligns the deformed mesh with the target point cloud, together with a mesh regularization loss that enforces smoothness and anatomically plausible surface reconstructions (\cite{park2025ai}).

Since all individual meshes were derived from the same template mesh, their vertex connectivity remains unchanged throughout the deformation process. As a result, the meshes preserve point-to-point vertex correspondence across subjects and time points. This consistent correspondence enables both cross-sectional and longitudinal statistical shape analyses at each mesh vertex. It is worth noting that, as the deformation is defined with respect to the template mesh, atrophy (inward deformation) is represented as a positive value. Therefore, as time progresses and the structure shrinks (see \autoref{fig:temp_mesh}), the deformation values increase.



\subsubsection{Computation of local deformity}

To analyze local shape variations, we calculate the local deformity of the reconstructed individual meshes. The local deformity is defined as a signed Euclidean distance function, following the formulation from a previous study~\cite{park2025ai}. But different from that cross-sectional study (\cite{park2025ai}), and due to the sequential nature of the process described previously, in this occasion, the signed Euclidean distance between a mesh point in the template and the corresponding point in the mesh at time \textit{t} is positive if the direction of the vector defining the displacement is towards the center of the template, and negative if the direction is outwards (i.e., opposite).

For cross-sectionally analyzing the shape variations with respect to the average shape of the sample at a reference time-point, we first constructed an average mesh by averaging the meshes of all subjects at the first scan (i.e., wave 2). We then computed the local deformity of each mesh-point for each subject with respect to this average mesh. These deformities were also used  to estimate overall trajectories of change across the 9-year period.

For the longitudinal analysis of shape deformations between two time-points, we derived the local deformity by directly comparing the vertex spatial positions of the individual meshes across time-points to calculate the displacements between the time-point pairs as illustrated in ~\autoref{fig:strategy}. Specifically, we computed the deformity between a mesh \( \mathrm{Mesh}_n \) at an initial \(n\)-th time-point and the follow-up mesh \( \mathrm{Mesh}_m \) at the \(m\)-th time-point, as the subtraction of the corresponding "baseline" mesh-points from the "follow-up" mesh-points. Thus, for shape deformations between waves 2 and 4, for example, the initial time point is the first scan (i.e., at wave 2), and the follow-up time-point is the third scan (i.e., at wave 4). It is worth noting that for this analysis, as we are not calculating the deformation from each individual time point to the template, but between two time-points, in the vertices where the structure shrinks the change takes negative values. If it would have been conceived otherwise, as the displacement vectors (template-to-individual mesh at the time-point \textit{t}) may have directly comparable directions but not spatial orientations, we would have needed to calculate the projections of these vectors in the straight line between their spatial locations at each time point, with a subsequent increase in processing time.



\subsection{Cognitive data processing}


The 13 cognitive tests applied at each time-point (i.e., testing wave) included five subtests from the Wechsler Adult Intelligence Scale (3rd edition): Block Design, Matrix Reasoning, Symbol Search, Digit-Symbol Substitution and Digit Span Backwards.  It also included three subtests from the Wechsler Adult Memory Scale (3rd edition): Verbal Paired Associates, Spatial Span and Logical Memory. The other five cognitive tests were: Verbal Fluency \cite{lezak2004neuropsychological}, Four-choice Reaction Time  \cite{deary2001reaction}, Inspection Time \cite{deary2004functional}, and two reading tests: the National Adult Reading Test (NART) \cite{nelson1991revised}, and the Weschler Test of Adult Reading (WTAR) \cite{wechsler2001wechsler}. To generate general cognitive functioning, or \textit{g}, scores, a latent growth curve model was calculated in a structural equation model framework with the lavaan package (v.0.6.17) \cite{rosseel2012lavaan} in R (v 4.2.0) \cite{team2020ra}. An intercept and slope was modeled for each of the 13 cognitive tests, and the 13 intercept factors loaded onto a g intercept factor, and the 13 slope factors loaded onto a g slope factor. Additional covariance paths were included at the wave-level between NART and WTAR, as these are similar reading tests.  NART had a small negative residual variance, which we fixed to 0 to allow the model to converge. The model had acceptable fit according to criteria provided by \citep{hu1999cutoff}): Comparative Fit Index (CFI) = 0.971 ($>$0.95), Tucker Lewis Index (TLI)= 0.970 ($>$0.95), Root Mean Square Error of Approximation (RMSEA)= 0.028 ($<$0.06), and Standardized Root Mean Squared Residual (SRMR)= 0.055 ($<$0.08). Missingness was modeled with full informational maximum likelihood to maximize statistical power, and slopes were removed for participants who only had one wave of data. We recovered estimated \textit{g} scores at each wave adding the extracted intercept value and the product of the slope and the time difference between both cognitive tests for a given participant.
Cognitive ability at age 11 was assessed using the Moray House Test IQ score from the Scottish Mental Survey of 1947.

\subsection{Statistical analyses}

We used the Matlab R2023a statistics and machine learning toolboxes and Python (Pytorch3d library) for statistical analyses. We used the Wilcoxon rank sum (Mann-Whitney U) test to compare the baseline clinical, demographic and cognitive characteristics of dropouts at each wave \textit{versus} those imaged, contributors to the analyses, and those who underwent cognitive testing. Repeated measures ANOVA was used to compare these characteristics in the samples that contributed to the longitudinal analyses. 

For reporting results of the shape variations/deformations in relation to general cognition, we follow the anatomical subregions reported in \cite{dekraker2020hippocampal} for the hippocampus, \cite{weeland2022thalamus} for the thalami and in \cite{amunts2005cytoarchitectonic} for the amygdala. For the rest of the structures, we report the shape surface variations in relation to the adjacent brain structures (e.g., brain ventricles, internal/external capsules) and features (e.g., locations of high likelihood of strings of enlarged perivascular spaces).

\subsubsection{Cross-sectional analyses}

In each wave separately, we evaluated the association between the individual variations at each mesh vertex with respect to the template shape and general cognition at the time point using general linear models. In these models, general cognition was the outcome variable, shape variations were the predictor, and childhood IQ was the covariate. 

\subsubsection{Longitudinal analyses}

We performed a 9-year interval analysis using linear mixed-effect models with participant as grouping variable, age 11 IQ as covariate, cross-sectional shape variations as predictor, and \textit{g} as outcome variable. We also used ANCOVA models to calculate the effect of the vertex-wise deformations between two time-points on the change in general cognition at 3-, 6- and 9-year intervals; i.e., between waves 2 and 3, 3 and 4, 4 and 5, 2 and 4, 3 and 5, and 2 and 5. 

\subsubsection{Cross-sectional and longitudinal analyses using additional covariates}

We repeated cross-sectional and two time-point-wise longitudinal analyses accounting, in addition, for brain tissue volume, sex, and vascular risk factors. We also repeated the analyses using the shape variations at each vertex adjusted by intracranial volume (ICV). The process of adjusting the shape morphology within-wave variations and between-wave changes by ICV follows the same principle as \cite{hernandez2017hippocampal}. Briefly, after the surface mesh templates generated as described above were fit to each subcortical binary mask, for each structure all meshes were co-registered and scaled using the individuals’ ICV, and an average mesh was generated again. This “ICV-adjusted" average mesh was then aligned with each individual mesh (i.e. one-by-one transformations to native space) to calculate the deformation of each point (i.e. mesh vertex) of each structure with respect to the correspondent point in this sample-specific ICV-adjusted shape. The ICV was not used in any model as covariate.

\backmatter


\bmhead{Acknowledgements}
 
Authors thank the LBC1936 Study participants and the wider team of radiographers, radiologists, geriatricians, and those that have been involved in the cognitive testing and image processing throughout the years, since 2004 until this date (2026).

\section*{Declarations}

\begin{itemize}
\item Funding

This work was supported by Institute for Information \& communications Technology Promotion(IITP) grant funded by the Korea government(MSIT) (No.00223446, Development of object-oriented synthetic data generation and evaluation methods), and the National Research Foundation of Korea(NRF) grant funded by the Korea government(MSIT) (RS-2024-00508681, Establishment of Korea-UK preclinical/clinical joint research center to develop diagnosis and treatment strategy for neurodegenerative diseases). The LBC1936 study was supported by Age UK (the Disconnected Mind project), the UK Medical Research Council (MRC; G0701120, G1001245, MR/M013111/1, MR/R024065/1), joint funding from the Medical Research Council and the Biotechnology and Biological Sciences Research Council (MR/K026992/1 for the Centre for Cognitive Ageing and Cognitive Epidemiology), joint funding from the Biotechnology and Biological Sciences Research Council and the Economic and Social Research Council (BB/W008793/1), and the University of Edinburgh. MR Imaging and the statistical analyses presented here were further supported by the Row Fogo Charitable Trust (The Row Fogo Centre for Research into Ageing and the Brain; AD.ROW4.35. BRO-D.FID3668413), and UK Dementia Research Institute (Edin002, DRIEdi17/18, and MRC MC-PC-17113) which receives its funding from DRI Ltd, funded by the UK Medical Research Council. SRC was also supported by a Sir Henry Dale Fellowship jointly funded by the Wellcome Trust and the Royal Society (221890/Z/20/Z).

\item Conflict of interest/Competing interests 

The authors declare that the work was conducted in the absence of any conflict of interest and that the funders did not play any role in the conception or in the analyses presented here.

\item Ethics approval and consent to participate

The LBC1936 study protocols were approved by the Multicentre Research Ethics Committee for Scotland (MREC/01/0/56), the Lothian Research Ethics Committee (LREC/2003/2/29), and the Scotland A Research Ethics Committee (07/MRE00/58). All LBC1936 study participants gave written consent to participate in the study and for their data to be used for research purposes. The participation in the study was entirely volunteer.

\item Consent for publication

All co-authors approved the manuscript to bue submitted for publication.

\item Data availability 

All image and cognitive data is available per request to the LBC1936 Study team (\url{https://lothian-birth-cohorts.ed.ac.uk/data-access-collaboration}). All the template models are freely available from Edinburgh DataShare (Link and DOI to be given. Currently under review).

\item Code availability 

The code, templates, vertex deformations, and results from all analyses are freely available from Edinburgh DataShare (Link and DOI to be given. Currently under review).

\item Author contribution

MCVH, WP: Conceptualization, Methodology, Software, Validation, Formal analysis, Investigation,  Data Curation, Writing - Original Draft, Visualization, Project administration

SMM, JM: Methodology, Software, Validation, Formal analysis, Investigation,  Visualization, Data Curation, Writing - Review and Editing, Visualization

JC, FNS: Data Curation, Writing - Review and Editing

MEB, JMW: Resources, Writing - Review and Editing, Funding acquisition

SRC: Conceptualization, Methodology,  Validation, Formal analysis, Investigation, Visualization, Resources,  Writing - Review and Editing, Supervision, Project administration, Funding acquisition

JP: Conceptualization, Methodology,  Validation, Resources,  Writing - Review and Editing, Supervision, Project administration, Funding acquisition

\end{itemize}







\begin{appendices}

\section{Methods - supporting figure and table}

\begin{figure}[H]
\centering
\includegraphics[width=1\textwidth]{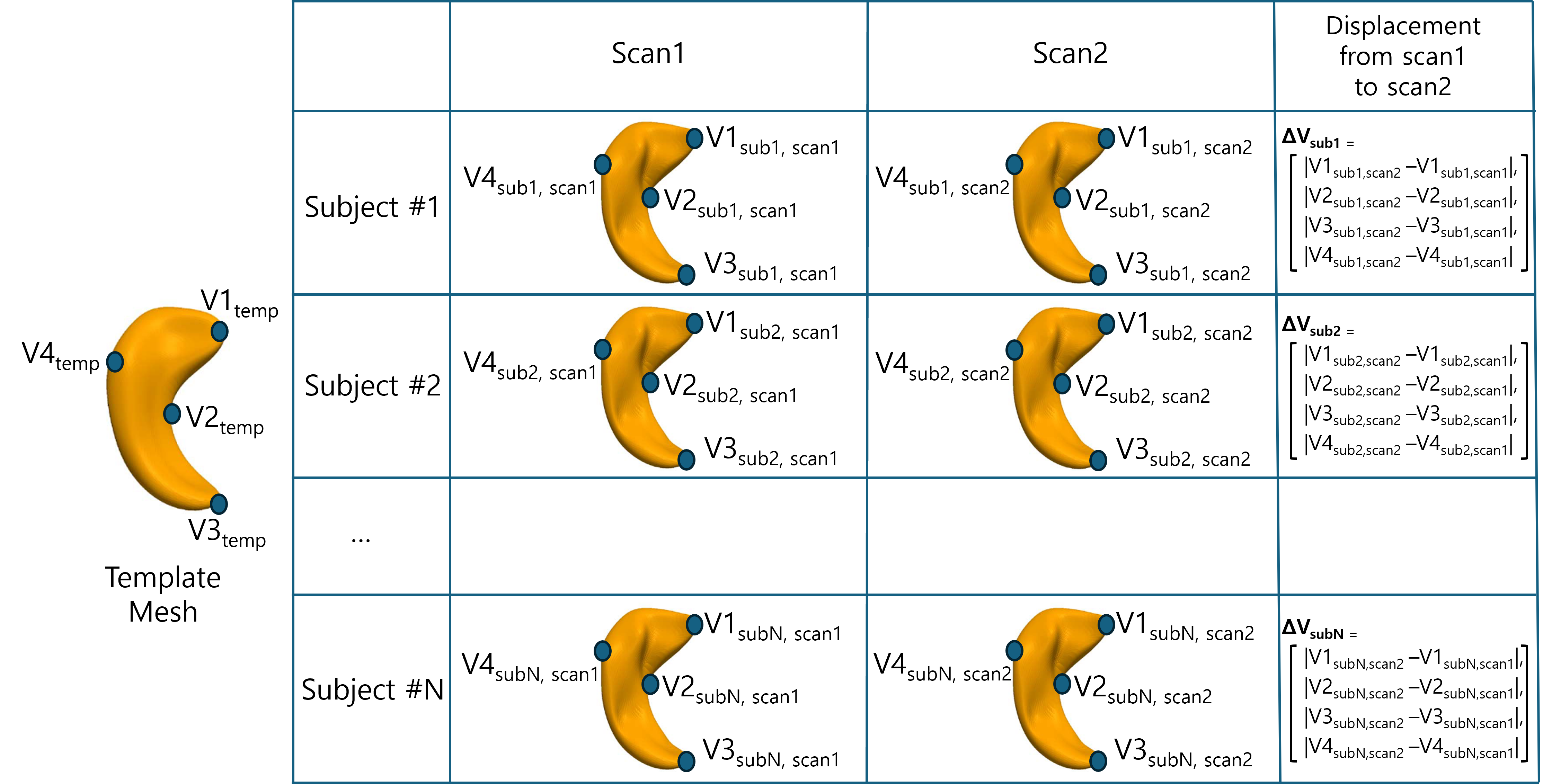}
\caption{Schematic illustration of the model data and the strategy for calculating the vertex displacement from scan 1 (wave 2) to scan 2 (wave 3) for the left hippocampus.}\label{fig:strategy}
\end{figure}

\begin{table}[h!]
\centering
\caption{Number of vertices of each shape template mesh}
\label{tab:number_of_vertices}
\begin{tabular}{lcc}
\toprule[.8mm]
\multicolumn{1}{c}{\multirow{2}{*}{Subcortical   structure}} & \multicolumn{2}{c}{Number of vertices} \\
\cmidrule{2-3}
\multicolumn{1}{c}{}                                         & Left Hemisphere   & Right Hemisphere   \\
\cmidrule{1-3}
Thalamus                                                     & 1456              & 1412               \\
Caudate                                                      & 1423              & 1492               \\
Putamen                                                      & 1549              & 1521               \\
Pallidum                                                     & 613               & 600                \\
Hippocampus                                                  & 1490              & 1492               \\
Amygdala                                                     & 623               & 631                \\
Accumbens                                                    & 336               & 325                \\
Ventral DC                                                   & 1307              & 1303               \\
\bottomrule[.8mm] 
\end{tabular}
\end{table}
\newpage
\section{Results - recruitment chart}

\begin{figure}[h]
\centering
\includegraphics[width=0.55\textwidth, height=0.8\textheight]
{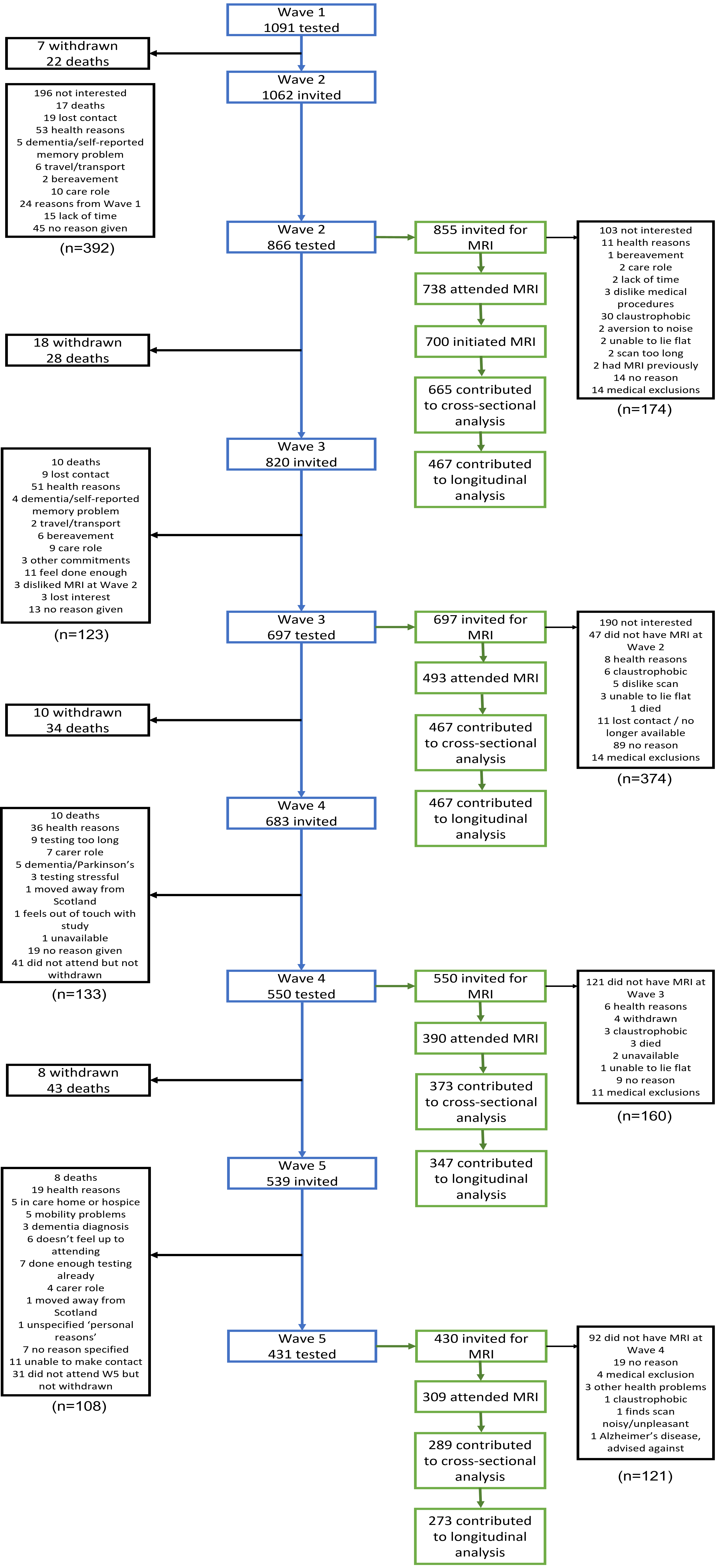}
\caption{Recruitment and attrition waves 1 to 5. }\label{fig:recruitment}
\end{figure}

\clearpage
\section{Results - Contributors to \textit{g}}

\begin{table}[h]
\centering
\caption{Sample size (N) and standardized intercept and slope loadings of the 13 cognitive tests on g (w2 to w5)}
\label{tab:tests_loadings_table}
\fontsize{8.2}{12}\selectfont
\setlength{\tabcolsep}{3.5pt}
\begin{tabular}{l *{9}{c}}
\toprule[.8mm]
                         & & \multicolumn{2}{c}{Intercept loadings} & & & \multicolumn{2}{c}{Slope loadings} & & \\
                         & N (w2, w3, w4, w5) & β & SE & z & p & β & SE & z & p \\
\midrule
Block design & 864, 691, 535, 420 & 0.664 & 0.031 & 21.475 & 0.000 & 0.964 & 0.161 & 6.002 & 0.000 \\
Matrix reasoning & 863, 689, 535, 418 & 0.749 & 0.029 & 25.389 & 0.000 & 0.688 & 0.131 & 5.254 & 0.000 \\
Spatial span & 861, 690, 536, 421 & 0.628 & 0.039 & 16.169 & 0.000 & 0.911 & 0.304 & 2.997 & 0.003 \\
Verbal fluency & 865, 696, 547, 426 & 0.519 & 0.033 & 15.950 & 0.000 & 0.725 & 0.086 & 8.422 & 0.000 \\
Verbal paired associates & 843, 663, 497, 380 & 0.557 & 0.034 & 16.371 & 0.000 & 0.659 & 0.057 & 11.472 & 0.000 \\
Logical memory & 864, 688, 542, 421 & 0.603 & 0.031 & 19.340 & 0.000 & 0.733 & 0.043 & 16.895 & 0.000 \\
Digit span backwards & 866, 695, 548, 426 & 0.672 & 0.031 & 21.917 & 0.000 & 0.379 & 0.116 & 3.272 & 0.001 \\
Symbol search & 862, 687, 528, 415 & 0.795 & 0.029 & 27.383 & 0.000 & 0.973 & 0.101 & 9.652 & 0.000 \\
Digit symbol & 862, 685, 535, 418 & 0.680 & 0.026 & 26.018 & 0.000 & 0.912 & 0.055 & 16.657 & 0.000 \\
Inspection time & 838, 654, 465, 382 & 0.494 & 0.039 & 12.556 & 0.000 & 0.868 & 0.083 & 10.502 & 0.000 \\
Choice reaction time & 865, 685, 543, 423 & 0.536 & 0.034 & 15.819 & 0.000 & 0.965 & 0.055 & 17.459 & 0.000 \\
NART & 864, 695, 546, 426 & 0.669 & 0.026 & 25.850 & 0.000 & 1.000 & 0.000 & NA & NA \\
WTAR & 864, 694, 546, 426 & 0.664 & 0.026 & 25.112 & 0.000 & 0.577 & 0.363 & 1.591 & 0.112 \\
\bottomrule[.8mm]
\end{tabular}
\end{table}

\section{Results - Missed values analysis}

\begin{longtable}[c]{lllll}
\caption{Results (i.e., p-values) from comparing the clinical, imaging, and cognitive data from participants who contributed to the analyses of change between testing waves \textit{versus} those who did not. For the interval between waves 2 to 3, 467 contributed, for the interval between waves 3 to 4, 347 contributed, and for the interval between waves 4 to 5, 273.}
\label{tab:TAB_Missing_values}\\
\toprule[.8mm]
            &                   & waves 2 to 3 & waves 3 to 4 & waves 4 to 5 \\
            \cmidrule{1-5}
\endfirsthead
\multicolumn{5}{c}%
{{\bfseries Table \thetable\ continued from previous page}} \\
\toprule[.8mm]
            &                   & waves 2 to 3 & waves 3 to 4 & waves 4 to 5 \\
            \cmidrule{1-5}
\endhead
\multirow{5}{*}{g}           & Tested vs Contrib & \textbf{0.0368}     & \textbf{0.000243}     & \textbf{0}            \\
            & Tested vs Missed    & \textbf{0.0204}     & \textbf{0.00498}     & \textbf{0.00117}     \\
            & Missed vs Contrib   & \textbf{0.000136}     & \textbf{0}            & \textbf{0}            \\
            & Imaged vs Contrib & 0.0767     & \textbf{0.000863}     & \textbf{0}            \\
            & Imaged  vs Missed   & \textbf{0.0147}     & \textbf{0.00358}     & \textbf{0.000866}     \\
            \cmidrule{1-5}
\multirow{5}{*}{age 11 IQ}     & Tested vs Contrib & 0.174     & 0.0529      & \textbf{0.00914}     \\
            & Tested vs Missed    & 0.132     & 0.138     & 0.121     \\
            & Missed vs Contrib   & \textbf{0.0131}     & \textbf{0.00309}     & \textbf{0.000317}     \\
            & Imaged vs Contrib & 0.186     & 0.0586     & \textbf{0.0115}     \\
            & Imaged vs Missed    & 0.159     & 0.167     & 0.153     \\
            \cmidrule{1-5}
\multirow{3}{*}{BHI}         & Missed vs Contrib   & \textbf{0.0110}     & \textbf{0.000127}     & \textbf{0}            \\
            & Imaged vs Contrib & 0.287     & \textbf{0.0313}     & \textbf{0.00135}     \\
            & Imaged vs Missed    & 0.0617      & \textbf{0.0230}     & \textbf{0}            \\
            \cmidrule{1-5}
\multirow{3}{*}{BPF}         & Missed vs Contrib   & \textbf{0.00976}     & \textbf{0.000345}     & \textbf{0}            \\
            & Imaged vs Contrib & 0.280     & \textbf{0.0451}      & \textbf{0.00388}      \\
            & Imaged vs Missed    & 0.0572     & \textbf{0.0332}     & \textbf{0}            \\
            \cmidrule{1-5}
\multirow{3}{*}{Sex}         & Missed vs Contrib   & 0.392     & 0.462       & \textbf{0}            \\
            & Imaged vs Contrib & 0.701     & 0.669     & 0.538     \\
            & Imaged vs Missed    & 0.545     & 0.674     & \textbf{0}            \\
            \cmidrule{1-5}
Atrophy     & Missed vs Contrib   & 0.895     & 0.450      & \textbf{0}            \\
deep        & Imaged vs Contrib & 0.954     & 0.665      & 0.346     \\
            & Imaged vs Missed    & 0.924     & 0.660     & \textbf{0}            \\
            \cmidrule{1-5}
Atrophy     & Missed vs Contrib   & 0.374     & 0.189     & 0.147     \\
superficial & Imaged vs Contrib & 0.698     & 0.451     & 0.310     \\
            & Imaged vs Missed    & 0.522     & 0.445     & 0.493     \\
            \cmidrule{1-5}
\multirow{3}{*}{High BP}     & Missed vs Contrib   & 0.915     & 0.160     & \textbf{0}            \\
            & Imaged vs Contrib & 0.962     & 0.413     & 0.338     \\
            & Imaged vs Missed    & 0.940     & 0.420     & \textbf{0}            \\
            \cmidrule{1-5}
\multirow{3}{*}{Diabetes}    & Missed vs Contrib   & 0.0604     & \textbf{0.0146}     & \textbf{0}            \\
            & Imaged vs Contrib & 0.389       & 0.139     & 0.126     \\
            & Imaged vs Missed    & 0.198     & 0.177     & \textbf{0}            \\
            \cmidrule{1-5}
High        & Missed vs Contrib   & 0.551     & 0.596     & \textbf{0}            \\
Cholesterol & Imaged vs Contrib & 0.790     & 0.758      & 0.987     \\
            & Imaged vs Missed    & 0.673     & 0.761      & \textbf{0}            \\
            \cmidrule{1-5}
CVD         & Missed vs Contrib   & 0.952     & 0.347     & \textbf{0.000009}     \\
History     & Imaged vs Contrib & 0.979     & 0.582     & 0.462      \\
            & Imaged vs Missed    & 0.966     & 0.592      & \textbf{0.000001} \\
\bottomrule[.8mm]            
\end{longtable}

\section{Results - vertex deformation in Linear Mixed Model}

\begin{figure}[H]
\centering
\includegraphics[width=0.8\textwidth]
{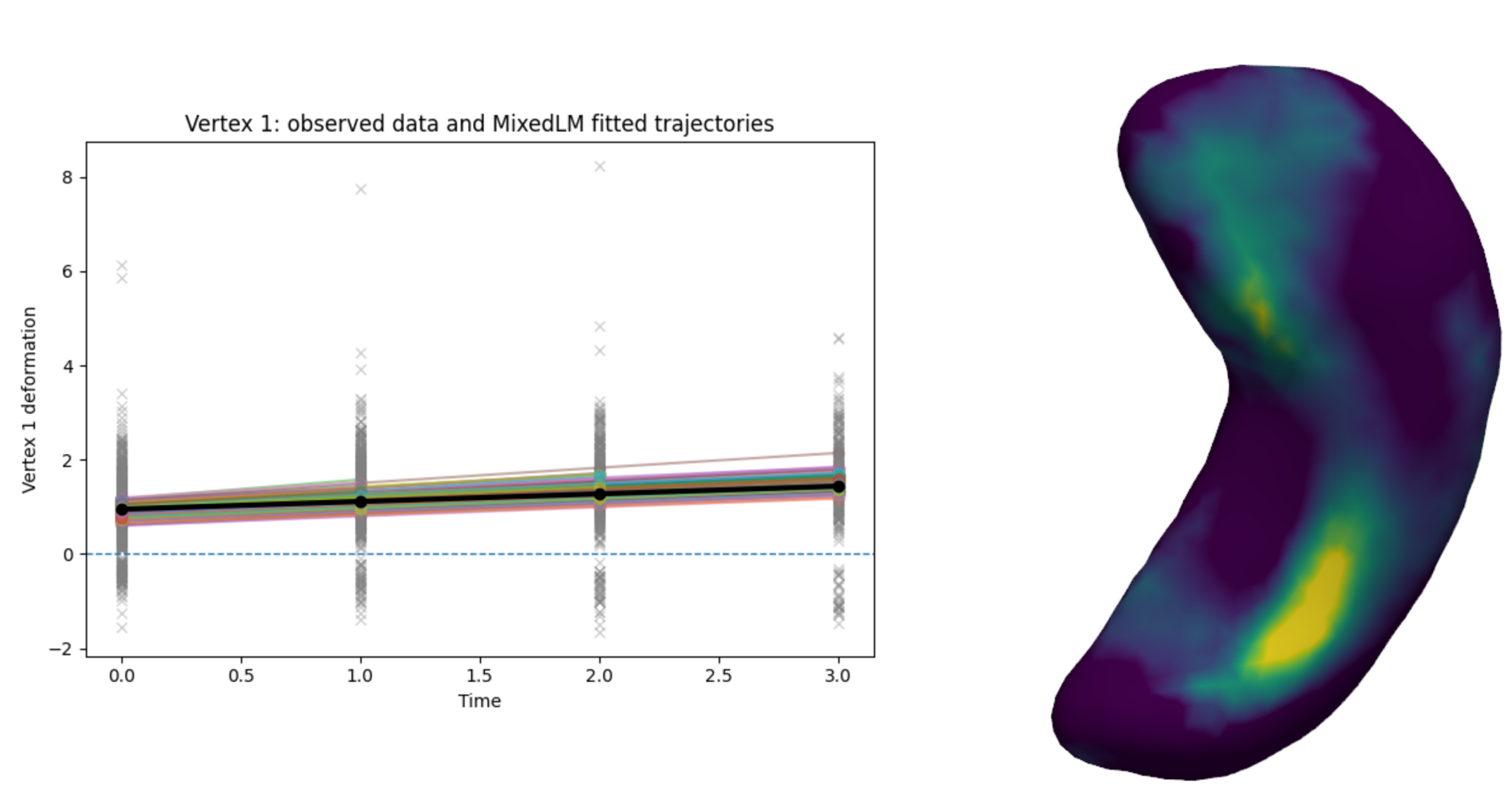}
\caption{Vertex deformation over the 9-year period (positive slope considering the cross-sectional measurements at the 4 time points) in a vertex where the linear mixed model yielded a negative association B=-0.184 (dark blue navy colour in the shape, p=6.35 x $10^{-05}$, SE=0.046, for the right hippocampus  }\label{fig:vertex_in_LMM}
\end{figure}

\clearpage
\section{Results - Linear Mixed Model, general results and correlations with the baseline associations}

\begin{figure}[H]
\centering
\includegraphics[width=1\textwidth]
{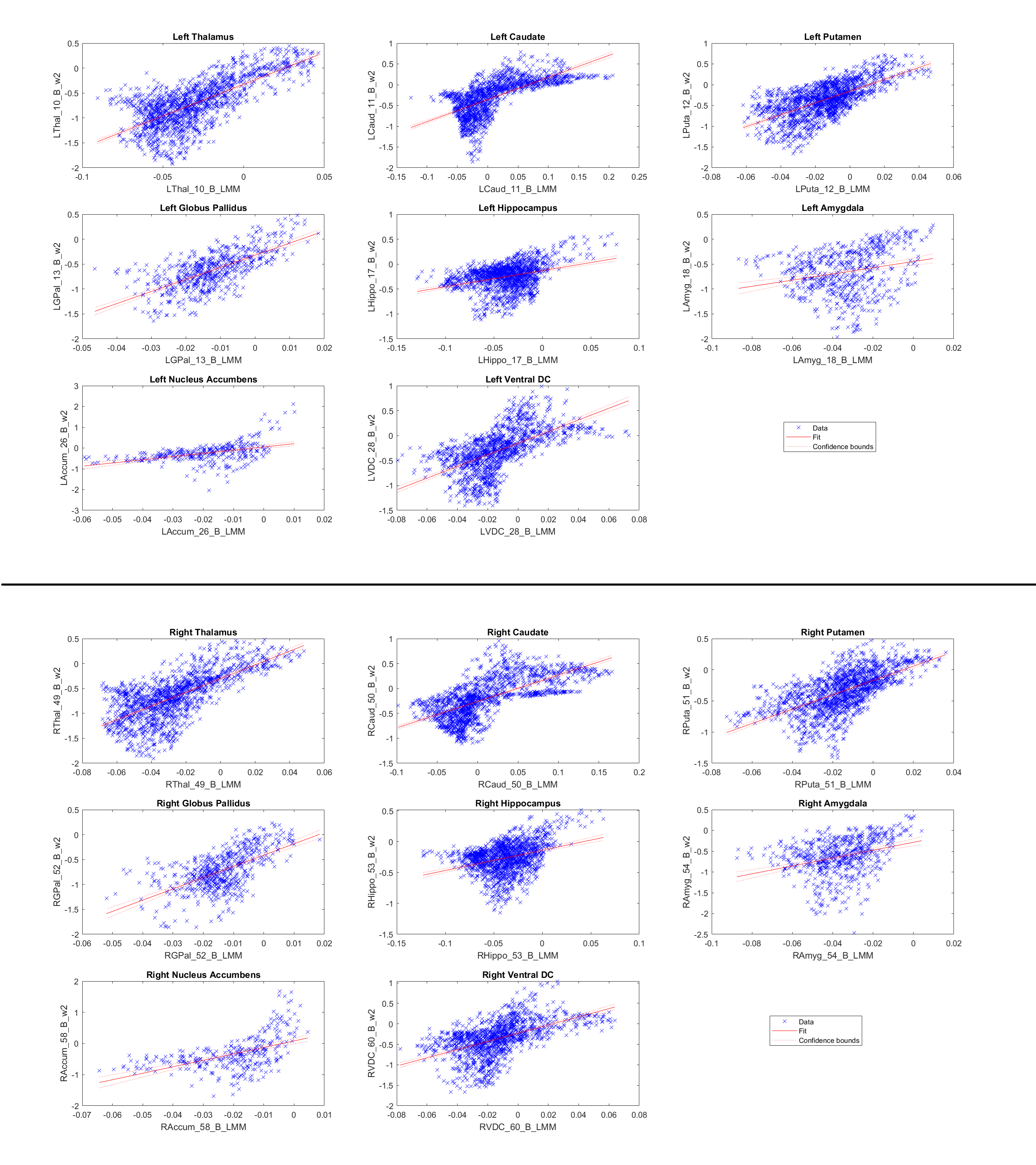}
\caption{Relationships between the average vertex-wise coefficient estimates b from the linear regression models (one per structure) applied at wave 2 (w2), and the linear mixed models (LMM) which analyze general change trajectories in cognition vs. shape deformations. Axes labels refer to the surfaces for each structure denoted as: brain hemisphere initial (i.e., L or R) followed by the 4-6 first letters of the structure's name, and a number that is the index label from SynthSeg.}\label{fig:correl_individual_struc}
\end{figure}


\begin{landscape}
\begin{table}[]
\caption{Correlations between the magnitude of the vertex-wise associations (coefficients estimates b) between general cognition and surface deformation patterns at the baseline point (i.e., wave 2, age 72.5 years, cross-sectional analysis) and the magnitude of the vertex-wise associations between changes in morphology and changes in general cognition throughout the nine years (i.e., results from the linear mixed models (LMM))}
\label{tab:correlations}
\begin{tabular}{llllllllllll}
\toprule[.8mm]
surface    & n\_vertices & Pearson\_r & spearman\_rho & cross\_mean & cross\_sd & long\_mean & long\_sd & mean\_abs\_cross & mean\_abs\_long & Cohen d & hemisphere \\
\cmidrule{1-12}
LAccum\_26 & 336         & 0.442983   & 0.476831    & -0.24275    & 0.46822   & -0.01917   & 0.013429 & 0.397395         & 0.01946         & 0.675 & L    \\
LAmyg\_18  & 623         & 0.239715   & 0.267782    & -0.65531    & 0.454621  & -0.03327   & 0.017532 & 0.668324         & 0.03347         & 1.9336 & L    \\
LCaud\_11  & 1423        & 0.659907   & 0.748057    & -0.29417    & 0.436115  & 0.011796   & 0.054077 & 0.399699         & 0.041671        & 0.9846 & L    \\
LGPal\_13  & 613         & 0.672755   & 0.663751    & -0.62606    & 0.374254  & -0.01253   & 0.010378 & 0.64654          & 0.013857        & 2.3175 & L    \\
LHippo\_17 & 1490        & 0.343635   & 0.285769    & -0.25297    & 0.25345   & -0.03749   & 0.02685  & 0.288499         & 0.039647        & 1.1957 & L    \\
LPuta\_12  & 1549        & 0.656869   & 0.642616    & -0.36568    & 0.406103  & -0.01486   & 0.018726 & 0.440356         & 0.019481        & 1.2204 & L    \\
LThal\_10  & 1456        & 0.712118   & 0.675255    & -0.70823    & 0.471629  & -0.03019   & 0.026387 & 0.736494         & 0.035407        & 2.030 & L     \\
LVDC\_28   & 1307        & 0.587225   & 0.678065    & -0.30909    & 0.434914  & -0.01337   & 0.021862 & 0.427812         & 0.021244        & 0.9604 & L    \\
RAccum\_58 & 325         & 0.513316   & 0.554132    & -0.33883    & 0.555206  & -0.02033   & 0.013715 & 0.552833         & 0.020385        & 0.7258 & R    \\
RAmyg\_54  & 631         & 0.323795   & 0.370865    & -0.63032    & 0.451127  & -0.03645   & 0.015406 & 0.641777         & 0.036465        & 0.7258 & R    \\
RCaud\_50  & 1492        & 0.664878   & 0.725844    & -0.19859    & 0.406472  & 0.011657   & 0.05106  & 0.375836         & 0.039611        & 0.7258 & R    \\
RGPal\_52  & 600         & 0.596863   & 0.576556    & -0.72924    & 0.390947  & -0.01412   & 0.010323 & 0.734773         & 0.014915        & 0.7258 & R    \\
RHippo\_53 & 1492        & 0.334619   & 0.308982    & -0.26223    & 0.241993  & -0.03803   & 0.024637 & 0.292541         & 0.039472        & 0.7258 & R    \\
RPuta\_51  & 1521        & 0.618416   & 0.661835    & -0.34808    & 0.32434   & -0.01534   & 0.017433 & 0.387787         & 0.018611        & 0.7258 & R    \\
RThal\_49  & 1412        & 0.690421   & 0.678487    & -0.67489    & 0.460828  & -0.02708   & 0.023067 & 0.702231         & 0.031325        & 1.9855 & R    \\
RVDC\_60   & 1303        & 0.517649   & 0.576217    & -0.36713    & 0.433357  & -0.01382   & 0.022441 & 0.450609         & 0.02176         & 0.7258 & R  \\
\bottomrule[.8mm] 
\end{tabular}
\caption*{\textit{Legend: Surfaces for each structure are labeled as follows: brain hemisphere initial (i.e., L or R) followed by the 4-6 first letters of the structure's name, and a number that is the index label from SynthSeg. Cross- and long- refer to the coefficient estimates (B-values) of the cross-sectional analysis at wave 2 and the longitudinal analysis using the LMM respectively. Absolute average coefficient estimate values are labeled as mean$\_$abs$\_$cross and mean$\_$abs$\_$long.}}
\end{table}
\end{landscape}

\section{Results - Associations between 3D shapes with size adjusted by ICV and cognition}\label{secA1}


\begin{figure}[H]
\centering
\includegraphics[width=1\textwidth]
{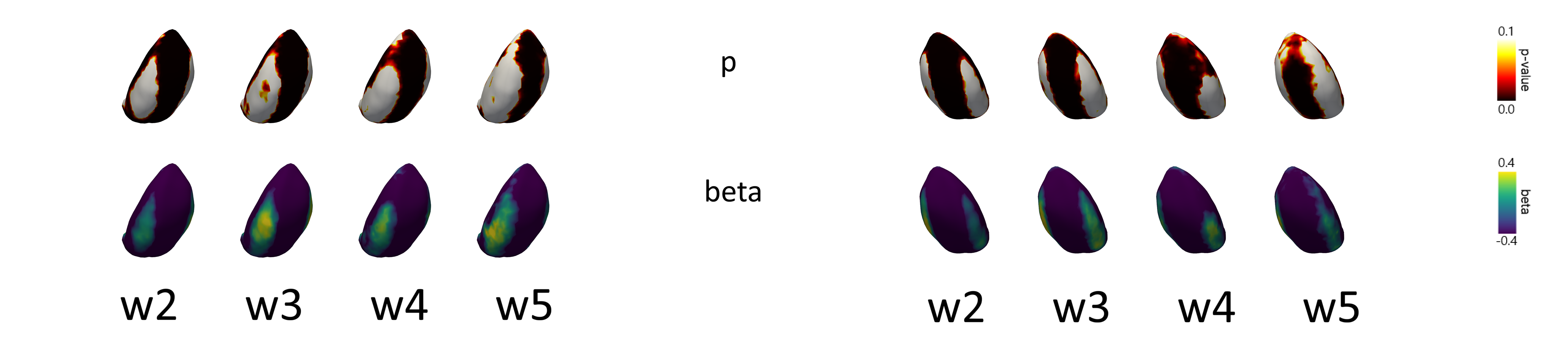}
    \begin{subfigure}[b]{0.48\textwidth}
        \centering
        \caption{L (ICV normalized)}
    \end{subfigure}
    \hfill
    \begin{subfigure}[b]{0.48\textwidth}
        \centering
        \caption{R (ICV normalized)}
    \end{subfigure}
\caption{Cross-sectional associations between thalamic shape adjusted by ICV and general cognition. From left to right, associations at waves 2, 3, 4 and 5.}\label{fig:icv_norm_thalami_cross}
\end{figure}
    

\begin{figure}[H]
\centering
\includegraphics[width=1\textwidth]
{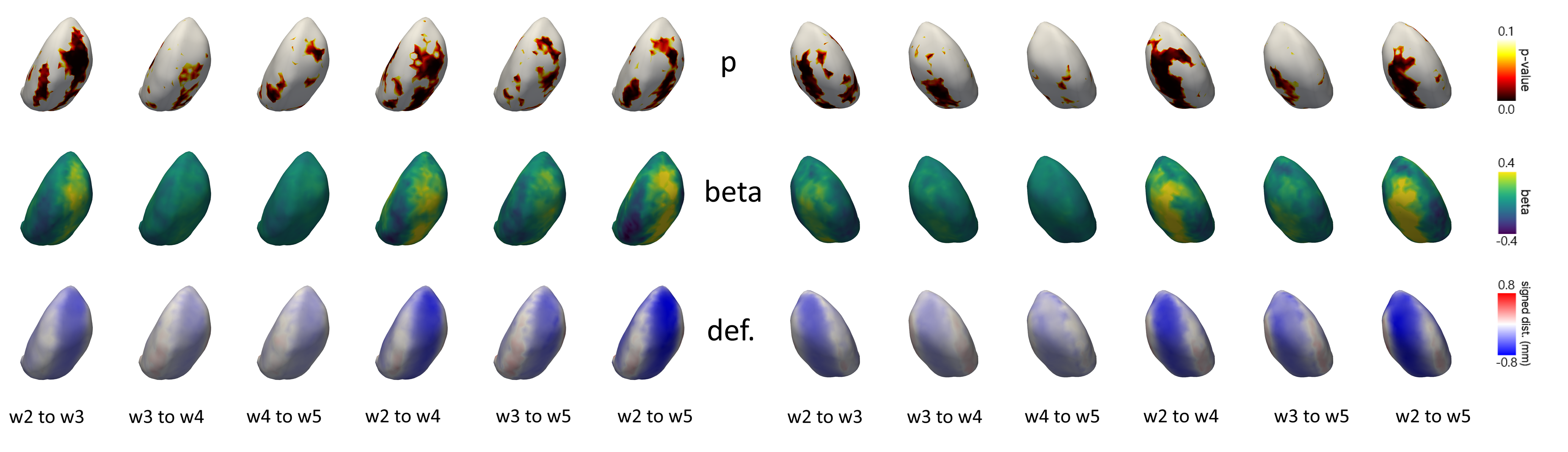}
    \begin{subfigure}[b]{0.48\textwidth}
        \centering
        \caption{L (ICV normalized)}
    \end{subfigure}
    \hfill
    \begin{subfigure}[b]{0.48\textwidth}
        \centering
        \caption{R (ICV normalized)}
    \end{subfigure}
\caption{Association of the shape changes of the left and right thalami (adjusted by ICV) and general cognitive changes, accounting for childhood intelligence, across different time intervals.}\label{fig:icv_norm_thalami_long}
\end{figure}


\begin{figure}[H]
\centering
\includegraphics[width=1\textwidth]
{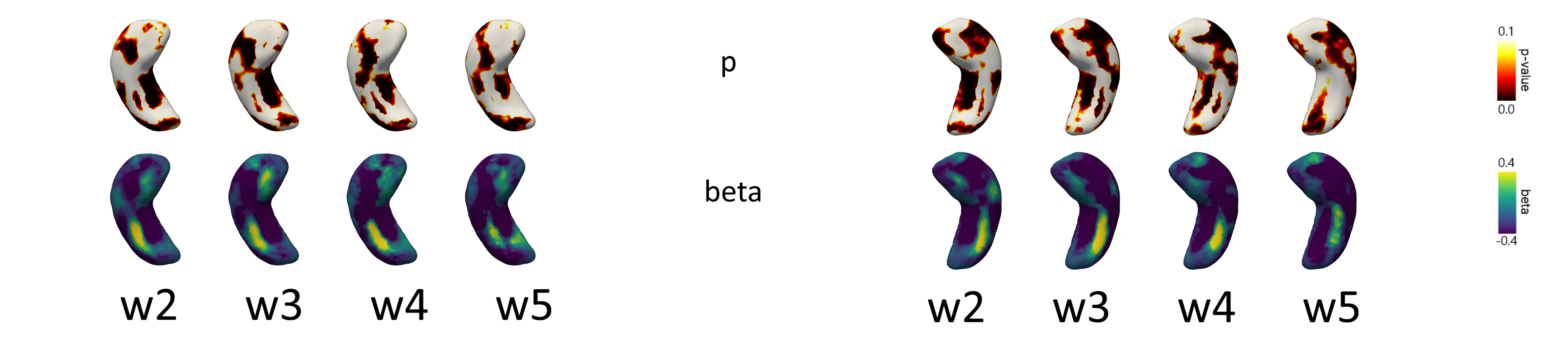}
    \begin{subfigure}[b]{0.48\textwidth}
        \centering
        \caption{L (ICV normalized)}
    \end{subfigure}
    \hfill
    \begin{subfigure}[b]{0.48\textwidth}
        \centering
        \caption{R (ICV normalized)}
    \end{subfigure}
\caption{Cross-sectional associations between hippocampal shape (adjusted by ICV) and cognition. From left to right, associations at waves 2, 3, 4 and 5.}\label{fig:icv_norm_hippocampi_cross}
\end{figure}


\begin{figure}[H]
\centering
\includegraphics[width=1\textwidth]
{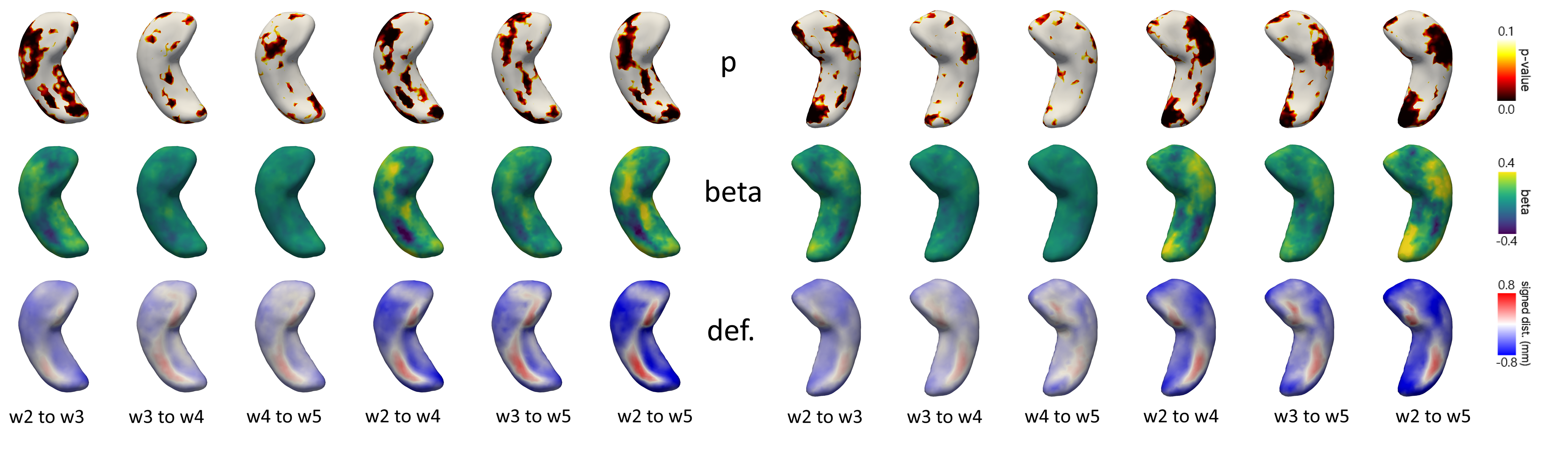}
    \begin{subfigure}[b]{0.48\textwidth}
        \centering
        \caption{L (ICV normalized)}
    \end{subfigure}
    \hfill
    \begin{subfigure}[b]{0.48\textwidth}
        \centering
        \caption{R (ICV normalized)}
    \end{subfigure}
\caption{Association of the shape of the left and right hippocampi (adjusted by ICV) and general cognition, accounting for childhood intelligence, across different time intervals.}\label{fig:icv_norm_hippocampi_long}
\end{figure}


\begin{figure}[H]
\centering
\includegraphics[width=1\textwidth]
{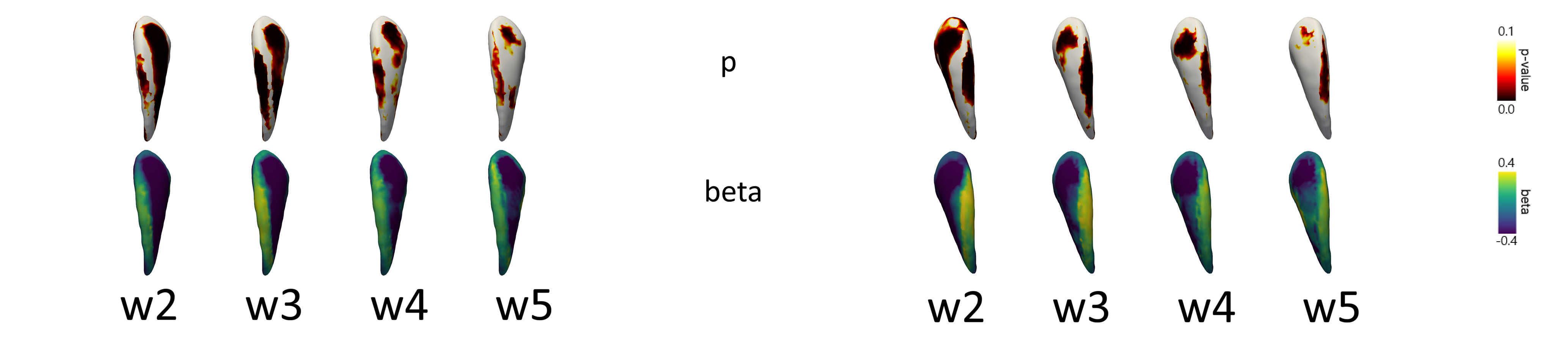}
    \begin{subfigure}[b]{0.48\textwidth}
        \centering
        \caption{L (ICV normalized)}
    \end{subfigure}
    \hfill
    \begin{subfigure}[b]{0.48\textwidth}
        \centering
        \caption{R (ICV normalized)}
    \end{subfigure}
\caption{Cross-sectional associations between caudate shape (adjusted by ICV) and cognition. From left to right, associations at waves 2, 3, 4 and 5.}\label{fig:icv_norm_caudate_cross}
\end{figure}


\begin{figure}[H]
\centering
\includegraphics[width=1\textwidth]
{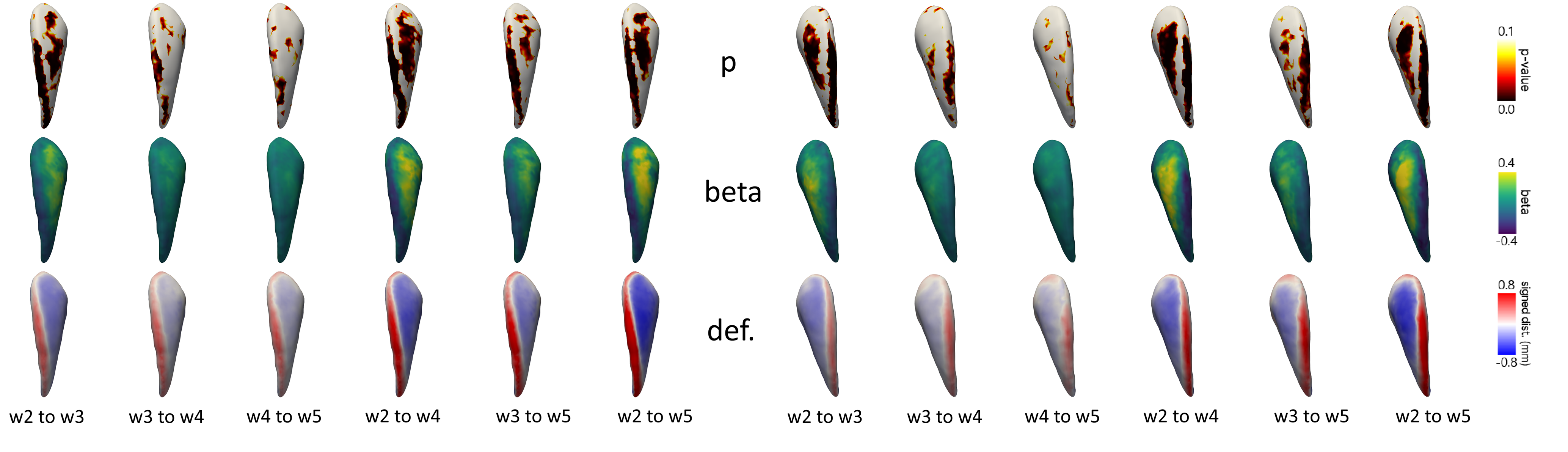}
    \begin{subfigure}[b]{0.48\textwidth}
        \centering
        \caption{L (ICV normalized)}
    \end{subfigure}
    \hfill
    \begin{subfigure}[b]{0.48\textwidth}
        \centering
        \caption{R (ICV normalized)}
    \end{subfigure}
\caption{Association of the shape of the left and right putamen (adjusted by ICV) and general cognition, accounting for childhood intelligence, across different time intervals.}\label{fig:icv_norm_caudate_long}
\end{figure}


\begin{figure}[H]
\centering
\includegraphics[width=1\textwidth]
{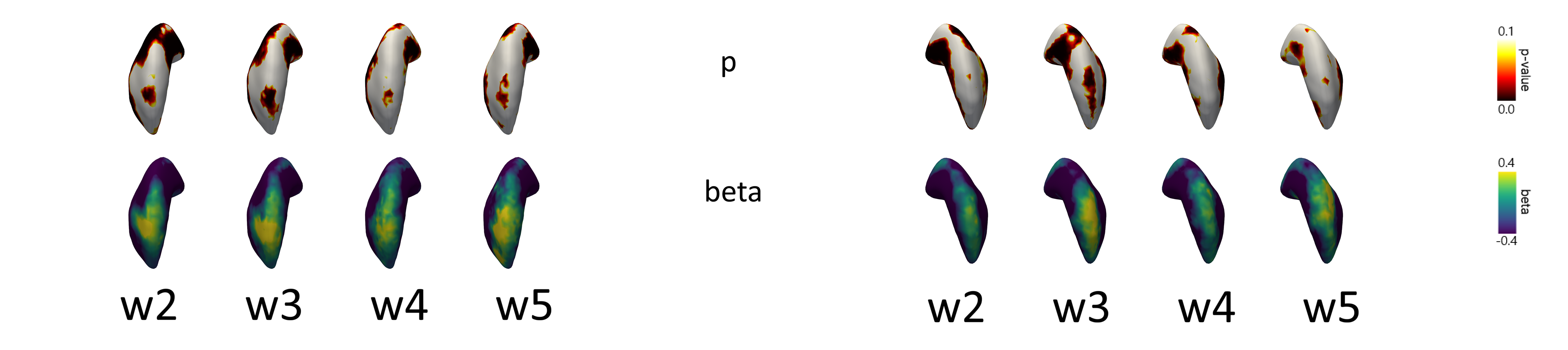}
    \begin{subfigure}[b]{0.48\textwidth}
        \centering
        \caption{L (ICV normalized)}
    \end{subfigure}
    \hfill
    \begin{subfigure}[b]{0.48\textwidth}
        \centering
        \caption{R (ICV normalized)}
    \end{subfigure}
\caption{Cross-sectional associations between putamen shape (adjusted by ICV) and cognition. From left to right, associations at waves 2, 3, 4 and 5.}\label{fig:icv_norm_putamen_cross}
\end{figure}


\begin{figure}[H]
\centering
\includegraphics[width=1\textwidth]
{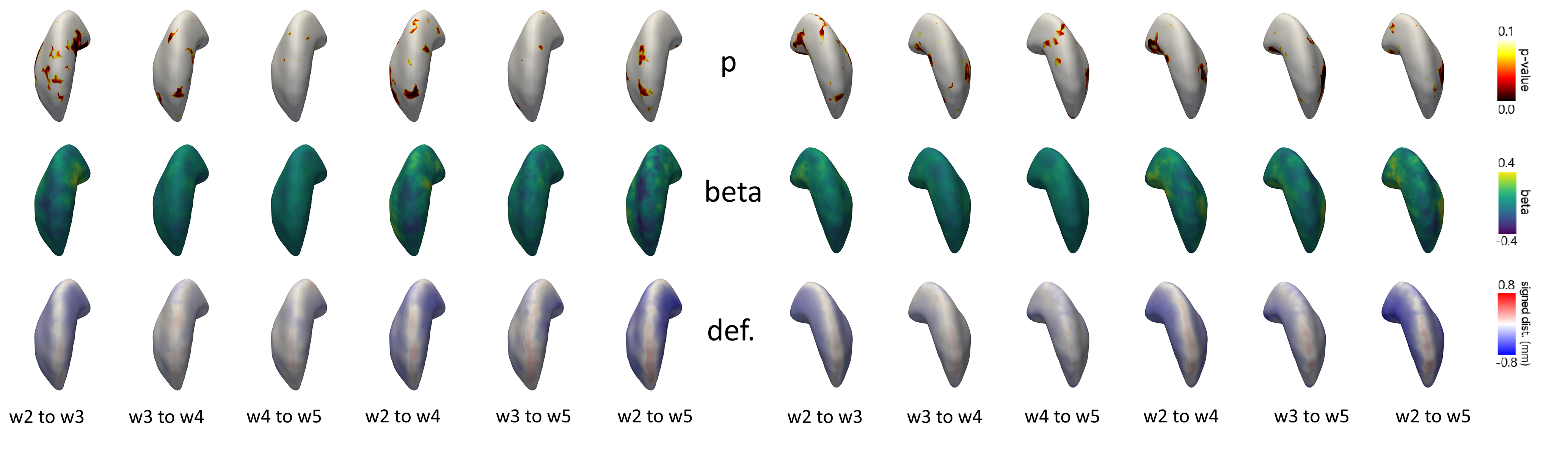}
    \begin{subfigure}[b]{0.48\textwidth}
        \centering
        \caption{L (ICV normalized)}
    \end{subfigure}
    \hfill
    \begin{subfigure}[b]{0.48\textwidth}
        \centering
        \caption{R (ICV normalized)}
    \end{subfigure}
\caption{Association of the shape of the left and right putamen (adjusted by ICV) and general cognition, accounting for childhood intelligence, across different time intervals.}\label{fig:icv_norm_putamen_long}
\end{figure}


\begin{figure}[H]
\centering
\includegraphics[width=1\textwidth]
{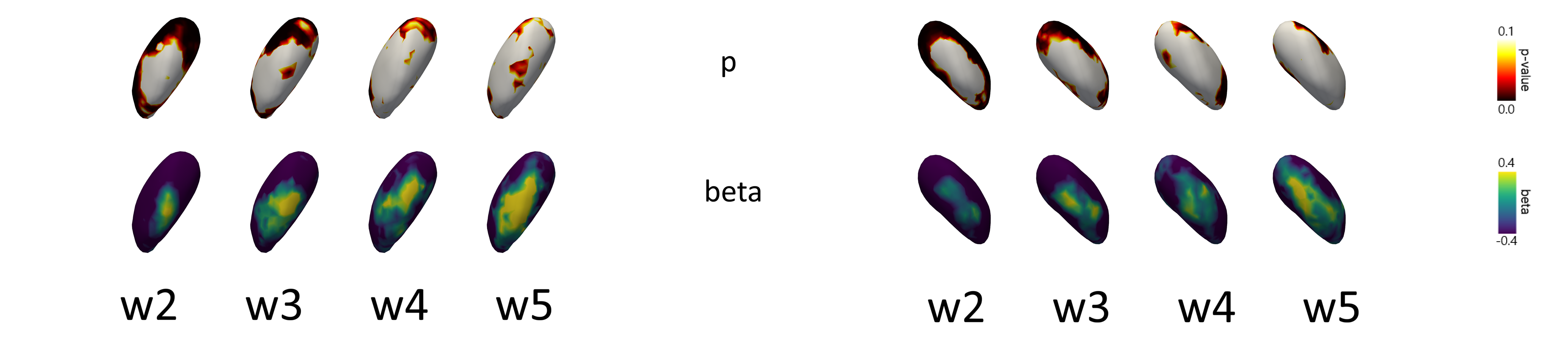}
    \begin{subfigure}[b]{0.48\textwidth}
        \centering
        \caption{L (ICV normalized)}
    \end{subfigure}
    \hfill
    \begin{subfigure}[b]{0.48\textwidth}
        \centering
        \caption{R (ICV normalized)}
    \end{subfigure}
\caption{Cross-sectional associations between the shape of the globus pallidus (adjusted by ICV) and cognition. From left to right, associations at waves 2, 3, 4 and 5.}\label{fig:icv_norm_gpallidi_cross}
\end{figure}


\begin{figure}[H]
\centering
\includegraphics[width=1\textwidth]
{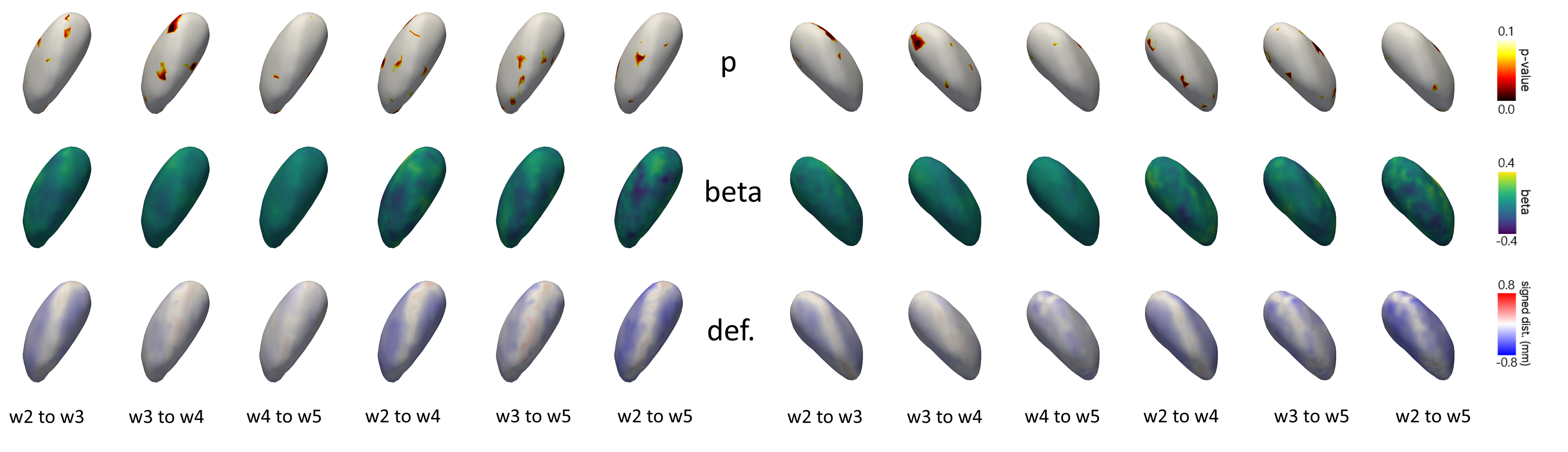}
    \begin{subfigure}[b]{0.48\textwidth}
        \centering
        \caption{L (ICV normalized)}
    \end{subfigure}
    \hfill
    \begin{subfigure}[b]{0.48\textwidth}
        \centering
        \caption{R (ICV normalized)}
    \end{subfigure}
\caption{Association of the shape of the left and right globus pallidi (adjusted by ICV) and general cognition, accounting for childhood intelligence, across different time intervals.}\label{fig:icv_norm_gpallidi_long}
\end{figure}

\begin{figure}[H]
\centering
\includegraphics[width=1\textwidth]
{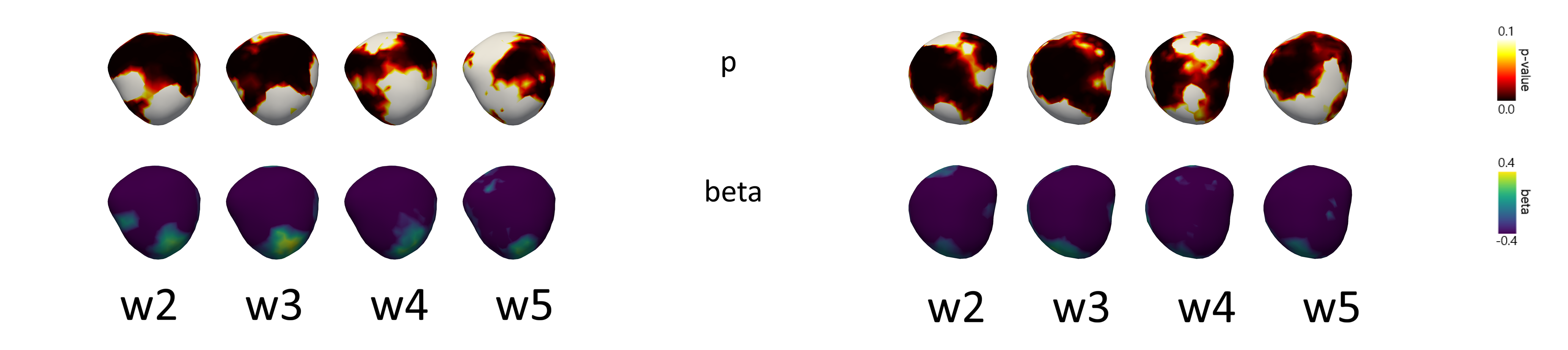}
    \begin{subfigure}[b]{0.48\textwidth}
        \centering
        \caption{L (ICV normalized)}
    \end{subfigure}
    \hfill
    \begin{subfigure}[b]{0.48\textwidth}
        \centering
        \caption{R (ICV normalized)}
    \end{subfigure}
\caption{Cross-sectional associations between the shape of the amygdala (adjusted by ICV) and cognition. From left to right, associations at waves 2, 3, 4 and 5.}\label{fig:icv_norm_Amygdala_cross}
\end{figure}


\begin{figure}[H]
\centering
\includegraphics[width=1\textwidth]
{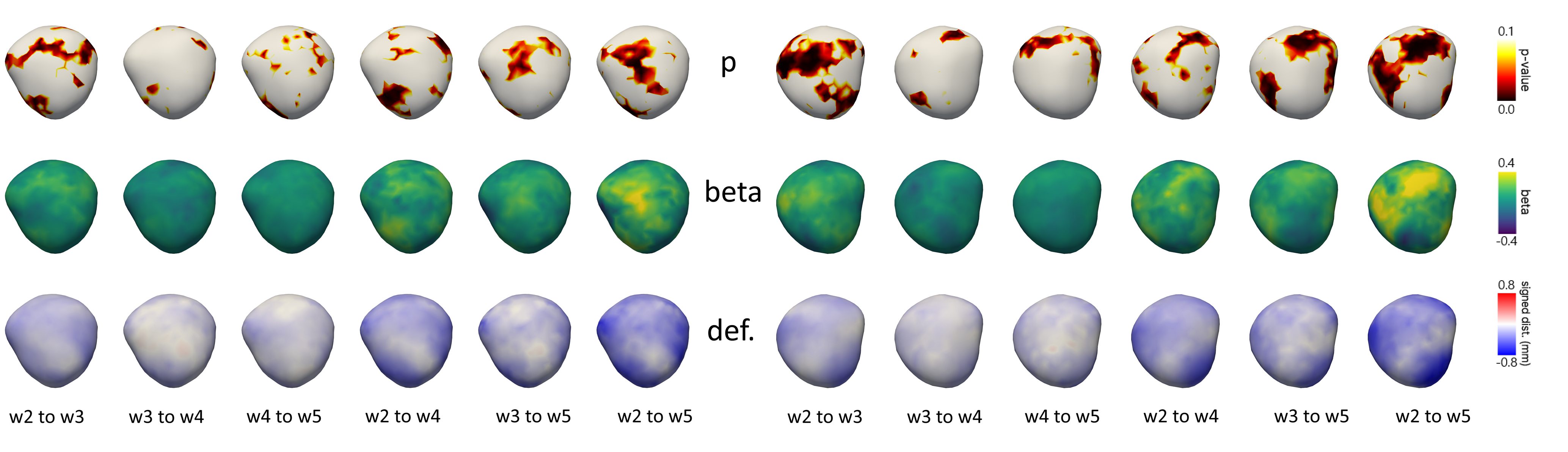}
    \begin{subfigure}[b]{0.48\textwidth}
        \centering
        \caption{L (ICV normalized)}
    \end{subfigure}
    \hfill
    \begin{subfigure}[b]{0.48\textwidth}
        \centering
        \caption{R (ICV normalized)}
    \end{subfigure}
\caption{Association of the shape of the left and right amygdala (adjusted by ICV) and general cognition, accounting for childhood intelligence, across different time intervals.}\label{fig:icv_norm_Amygdala_long}
\end{figure}


\begin{figure}[H]
\centering
\includegraphics[width=1\textwidth]
{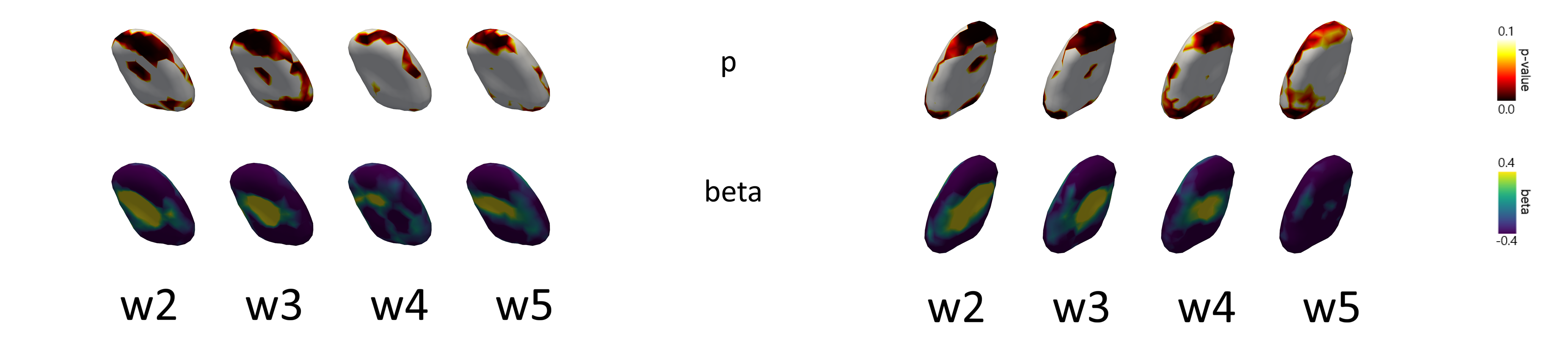}
    \begin{subfigure}[b]{0.48\textwidth}
        \centering
        \caption{L (ICV normalized)}
    \end{subfigure}
    \hfill
    \begin{subfigure}[b]{0.48\textwidth}
        \centering
        \caption{R (ICV normalized)}
    \end{subfigure}
\caption{Cross-sectional associations between the shape of the nucleus accumbens (adjusted by ICV) and cognition. From left to right, associations at waves 2, 3, 4 and 5.}\label{fig:icv_norm_Accumbens_cross}
\end{figure}


\begin{figure}[H]
\centering
\includegraphics[width=1\textwidth]
{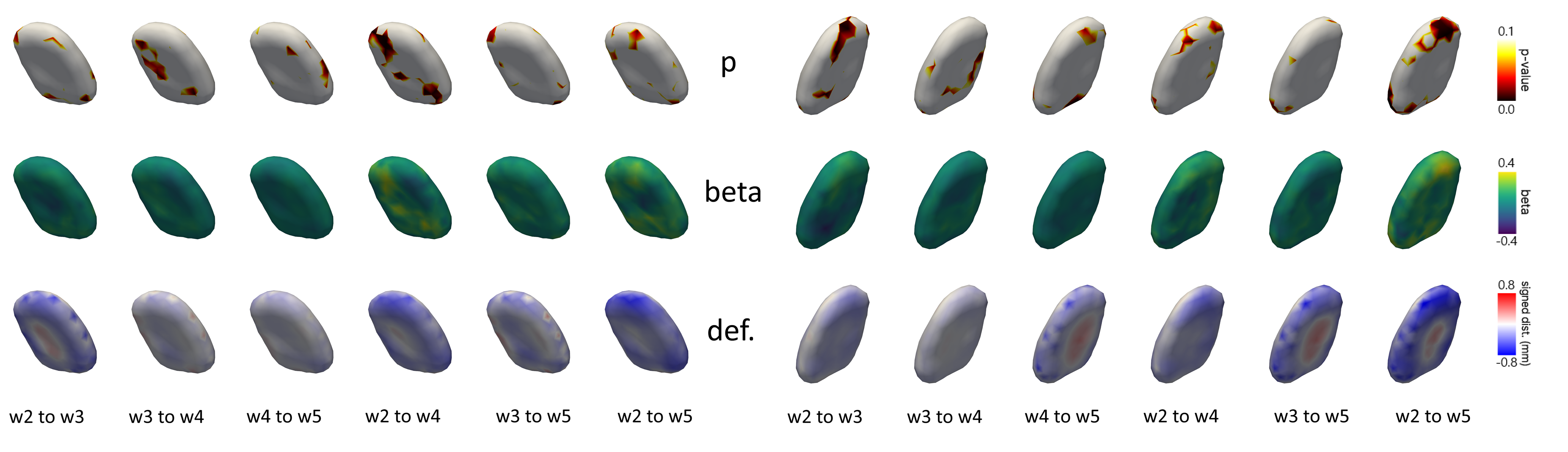}
    \begin{subfigure}[b]{0.48\textwidth}
        \centering
        \caption{L (ICV normalized)}
    \end{subfigure}
    \hfill
    \begin{subfigure}[b]{0.48\textwidth}
        \centering
        \caption{R (ICV normalized)}
    \end{subfigure}
\caption{Association of the shape of the left and right nucleus accumbens (adjusted by ICV) and general cognition, accounting for childhood intelligence, across different time intervals.}\label{fig:icv_norm_Accumbens_long}
\end{figure}


\begin{figure}[H]
\centering
\includegraphics[width=1\textwidth]
{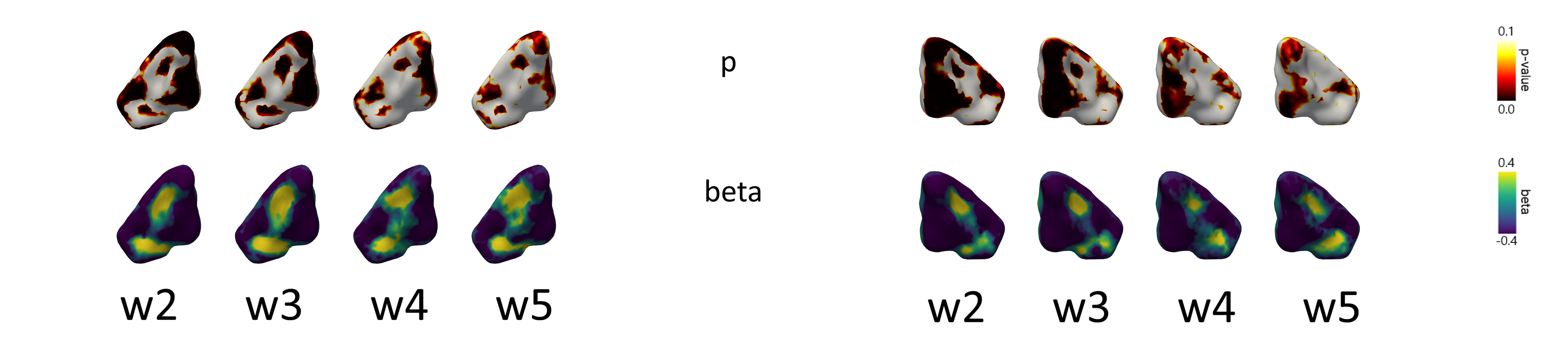}
    \begin{subfigure}[b]{0.48\textwidth}
        \centering
        \caption{L (ICV normalized)}
    \end{subfigure}
    \hfill
    \begin{subfigure}[b]{0.48\textwidth}
        \centering
        \caption{R (ICV normalized)}
    \end{subfigure}
\caption{Cross-sectional associations between the shape of the ventral diencephalon (adjusted by ICV) and cognition. From left to right, associations at waves 2, 3, 4 and 5.}\label{fig:icv_norm_VentralDC_cross}
\end{figure}


\begin{figure}[H]
\centering
\includegraphics[width=1\textwidth]
{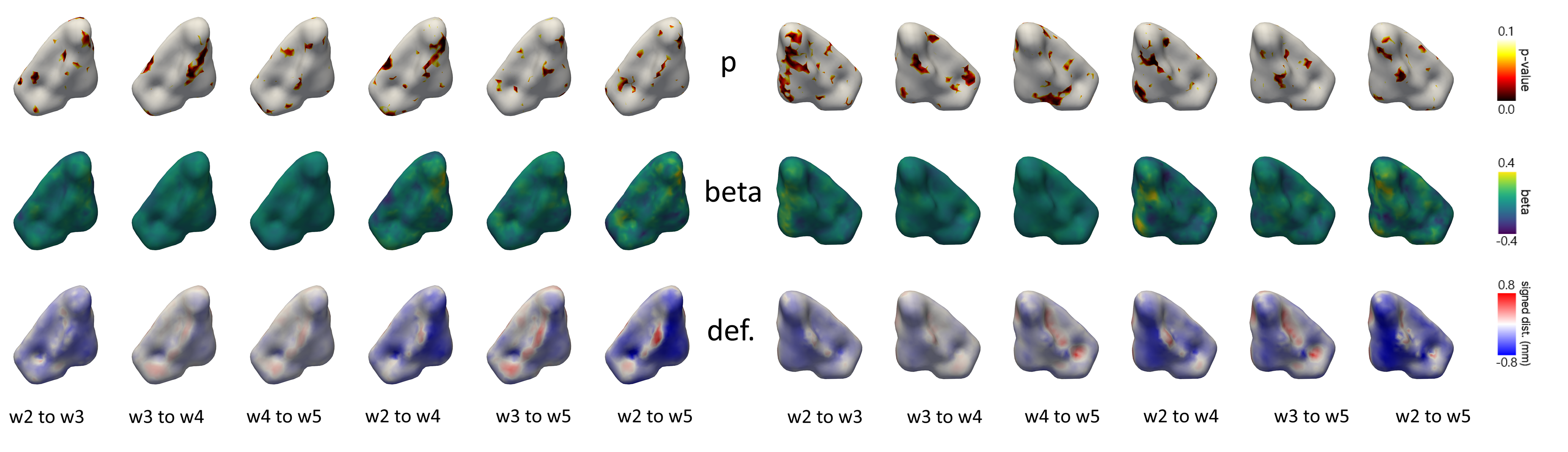}
    \begin{subfigure}[b]{0.48\textwidth}
        \centering
        \caption{L (ICV normalized)}
    \end{subfigure}
    \hfill
    \begin{subfigure}[b]{0.48\textwidth}
        \centering
        \caption{R (ICV normalized)}
    \end{subfigure}
\caption{Association of the shape of the left and right ventral diencephalon (adjusted by ICV) and general cognition, accounting for childhood intelligence, across different time intervals.}\label{fig:icv_norm_VentralDC_long}
\end{figure}


\section{Results - Sensitivity analysis}

\begin{figure}[H]
\centering
\includegraphics[width=1\textwidth]
{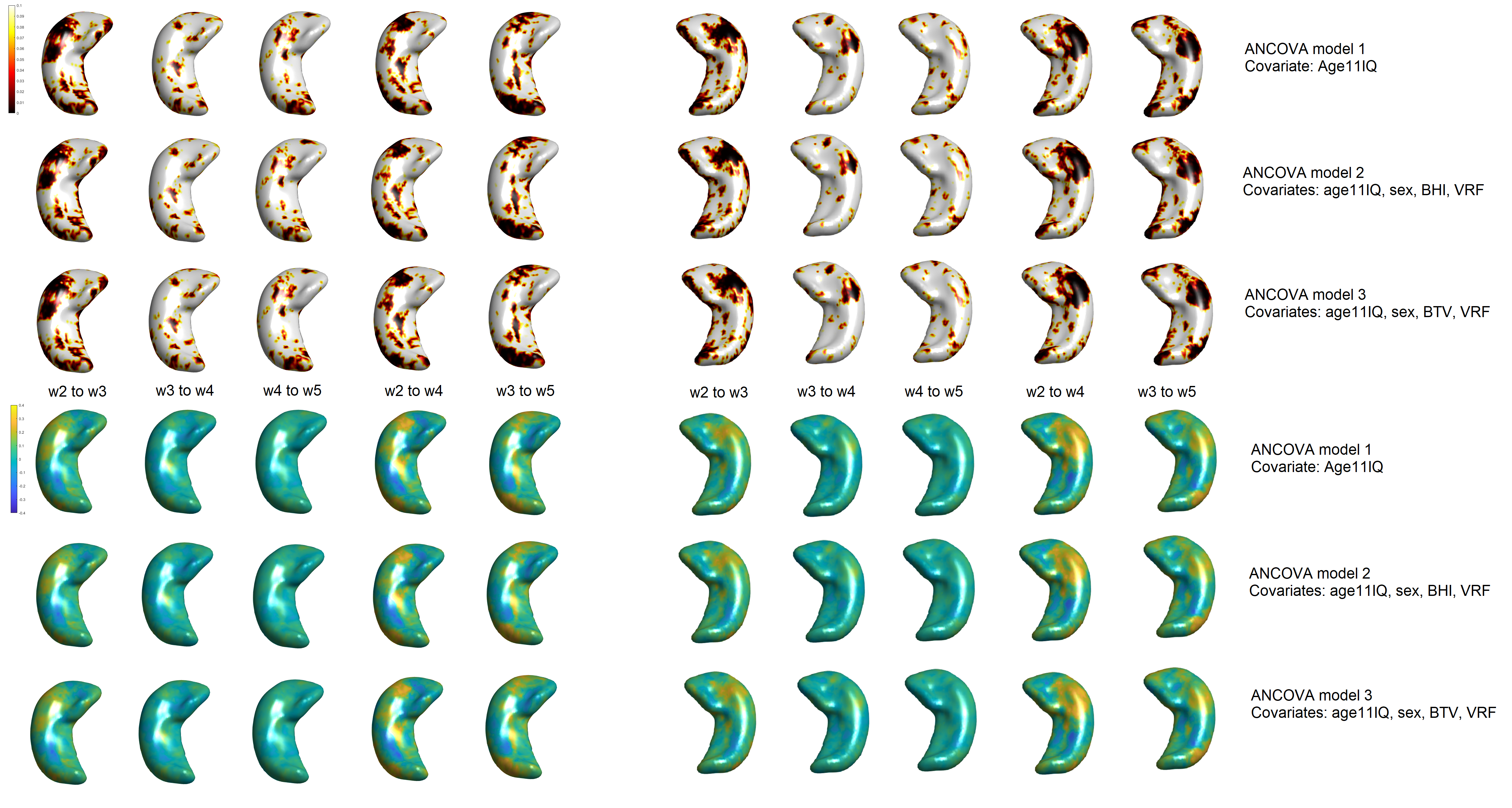}
\caption{Sensitivity analysis. Associations between \textit{g} and hippocampal shape deformations in different time-intervals. Statistical significance (above), and B-values (below).}\label{fig:sensitivity_long}
\end{figure}

\begin{figure}[H]
\centering
\includegraphics[width=1\textwidth]
{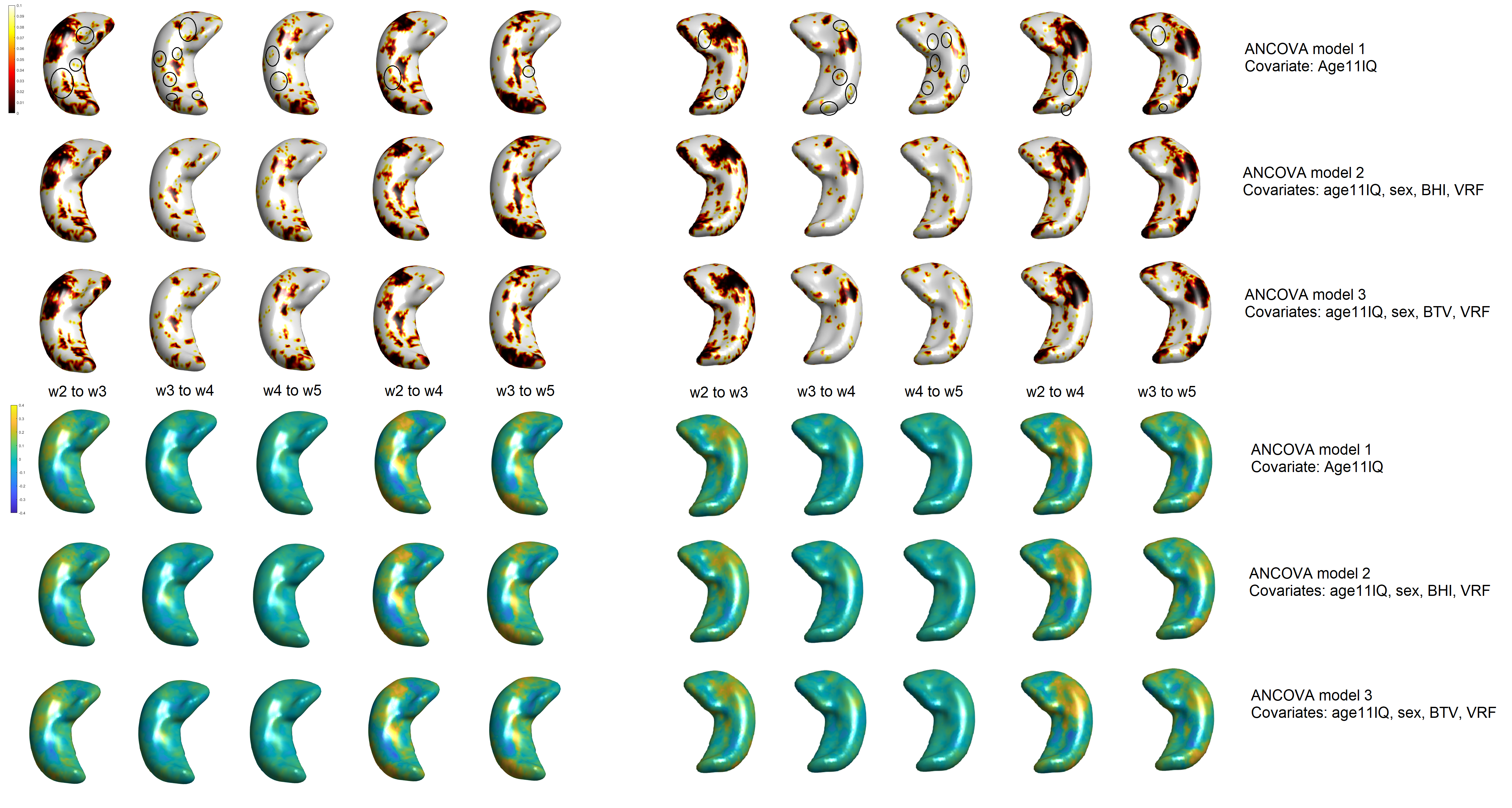}
\caption{Sensitivity analysis. Associations between \textit{g} and hippocampal shape deformations in different time intervals using shapes adjusted by ICV. Areas where differences between the outcome from the different statistical models were observed are encircled. Statistical significance (above), and B-values (below).}\label{fig:sensitivity_long_ICV}
\end{figure}




\end{appendices}


\bibliography{sn-bibliography}

\end{document}